\documentstyle[12pt,a41,epsfig,axodraw]{article}
\newenvironment{Feynman}[3]{\begin{center}
                            \setlength{\unitlength}{#3 mm}
                            \begin{picture}(#1)(#2)
                            \thicklines
                           }{\end{picture} \end{center}}

\newcommand{\be}{\begin{equation}}
\newcommand{\ee}{\end{equation}}
\newcommand{\bq}{\begin{equation}}
\newcommand{\eq}{\end{equation}}

\newcommand\GeV{\,\mbox{GeV}}

\newcommand{\ba}{\begin{eqnarray}}
\newcommand{\ea}{\end{eqnarray}}

\newcommand{\vph}{ \vphantom{\int\limits_t^t} }

\newcommand{\mr}{\rm}
\newcommand{\ds}{\displaystyle}

\newcommand{\nll}{\nonumber \\}

\newcommand {\Ql  }{\mbox{$Q^2_{l}    $}}
\newcommand {\QlS }{\mbox{$Q^4_{l}    $}}
\newcommand {\Qh  }{\mbox{$Q^2_{h}    $}}
\newcommand {\Qlh }{\mbox{$Q^2_{lh}   $}}
\newcommand {\QhS }{\mbox{$Q^4_{h}    $}}

\newcommand {\yl  }{\mbox{$y  _{l}    $}}
\newcommand {\ylS }{\mbox{$y^2_{l}    $}}
\newcommand {\xl  }{\mbox{$x  _{l}    $}}
\newcommand {\yll }{\mbox{$y  _{l_1}  $}}
\newcommand {\yllS}{\mbox{$y^2_{l_1}  $}}
\newcommand {\yh  }{\mbox{$y_{h}      $}}
\newcommand {\yhS }{\mbox{$y^2_{h}    $}}
\newcommand {\xh  }{\mbox{$x  _{h}    $}}
\newcommand {\ylh }{\mbox{$y  _{lh}   $}}
\newcommand {\ylhS}{\mbox{$y^2_{lh}   $}}
\newcommand {\ylpl}{\mbox{$Y  _{+}    $}}
\newcommand {\yhl }{\mbox{$y_{h_1}    $}}

\newcommand {\Me  }{\mbox{$m^2        $}}
\newcommand {\Mp  }{\mbox{$M^2        $}}
\newcommand {\SLS }{\mbox{$\sqrt{\lambda_S}  $}}
\newcommand {\SLQ }{\mbox{$\sqrt{\lambda_q}  $}}
\newcommand {\SLQS}{\mbox{$\lambda_q         $}}
\newcommand{\ClS  }{\mbox{$ \sqrt{C_1}                 $}}
\newcommand{\CllS }{\mbox{$ \sqrt{C_2}                 $}}
\newcommand{\lClS }{\mbox{${\ds \frac{1}{\sqrt{C_1}}}  $}}
\newcommand{\lCllS}{\mbox{${\ds \frac{1}{\sqrt{C_2}}}  $}}
\newcommand{\lCl  }{\mbox{${\ds \frac{1}{{C_1}^{3/2}}} $}}
\newcommand{\lCll }{\mbox{${\ds \frac{1}{{C_2}^{3/2}}} $}}
\newcommand{\ClSS }{\mbox{$      {C_1}^{3/2}           $}}

\newcommand{\BZl  }{\mbox{$B_1                         $}}
\newcommand{\BCl  }{\mbox{${\ds\frac{B_1}{{C_1}^{3/2}}}$}}
\newcommand{\BCll }{\mbox{${\ds\frac{B_2}{{C_2}^{3/2}}}$}}
\newcommand{\FCZl }
                  {\mbox{${\ds \frac{1}
                  {\sqrt{C_1}\left(B_1+\sqrt{\lambda_q}
                   \sqrt{C_1}\right)}} $}             }
\newcommand{\FCZll}
                  {\mbox{${\ds \frac{1}
                  {\sqrt{C_2}\left(B_2+\sqrt{\lambda_q}
                   \sqrt{C_2}\right)}} $}             }
\newcommand{\yllFCZll}
                     {\mbox{${\ds \frac{\yll}
                     {\sqrt{C_2}\left(B_2+\sqrt{\lambda_q}
                      \sqrt{C_2}\right)}} $}             }
\newcommand{\FCZ  }
                  {\mbox{${\ds  \frac{1}{\Qlh}
                   \left( \frac{1}{\sqrt{C_1}\left(B_1+\sqrt{\lambda_q}
                   \sqrt{C_1}\right)}
                  -\frac{1}{\sqrt{C_2}\left(B_2+\sqrt{\lambda_q}
                   \sqrt{C_2}\right)}\right)}$}}

\newcommand {\ZlS  }{\mbox{$z^2_{1} $}}

\newcommand {\Qe   }{\mbox{$Q_{e}   $}}

\newcommand {\COSPHI }{\mbox{$\cos{\varphi}                $}}

\newcommand {\SQSK   }{\mbox{$\lambda_{kqh}                $}}
\newcommand {\SQSKS  }{\mbox{$\sqrt {\lambda_{kqh}}        $}}
\newcommand {\SQSll  }{\mbox{$\sqrt{\lambda_{Slq}}         $}}

\newcommand{\Pe   }{\mbox{$p_l      $}}
\newcommand{\pe   }{\mbox{$p_l      $}}

\newcommand{\BZ    }{\mbox{$ B^2    _{1(2)}    $}}
\newcommand{\CZ    }{\mbox{$ C      _{1(2)}    $}}
\newcommand{\SC    }{\mbox{$ C^{1/2}_{1(2)}    $}}

\newcommand{\LS    }{\mbox{$     {\lambda_S}   $}}
\newcommand{\SLL   }{\mbox{$\sqrt{\lambda_l}   $}}
\newcommand{\LL    }{\mbox{$     {\lambda_l}   $}}

\newcommand{\LH    }{\mbox{$     {\lambda_h}   $}}
\newcommand{\SLK   }{\mbox{$\sqrt{\lambda_k}   $}}
\newcommand{\LK    }{\mbox{$     {\lambda_k}   $}}
\newcommand{\LQ    }{\mbox{$     {\lambda_q}   $}}

\newcommand{\Ncg}{ \mbox{${\LS-\LQ-\LL} $}}
\newcommand{\Nck}{ \mbox{${\LK+\LQ-\LH} $}}
\newcommand{\NSg}{ \mbox{${\sqrt{-\lambda(\LS,\LL,\LQ) }}$}}
\newcommand{\NSk}{ \mbox{${\sqrt{-\lambda(\LK,\LL,\LH) }}$}}
\begin{document}
\sloppy
\thispagestyle{empty}
\begin{flushleft}
DESY 96--189 \\
{\tt hep-ph/9612435}\\
December  1996 \\
\end{flushleft}

\mbox{}
\vspace*{\fill}
\begin{center}
{\LARGE\bf
\mbox{\boldmath $O(\alpha)$} QED Corrections to Neutral Current} \\

\vspace{1mm}
{\LARGE\bf Polarized Deep--Inelastic}\\

\vspace{1mm}
{\LARGE\bf Lepton--Nucleon Scattering}\\

\vspace{2em}
\large
Dmitri Bardin$^{a,b}$,
Johannes Bl\"umlein$^a$,
Pena Christova$^{a,c}$,\\
and
Lida Kalinovskaya$^{a,b}$
\\
\vspace{2em}
\normalsize
{\it $^a$DESY--Zeuthen,}
 \\
{\it Platanenallee 6,
D--15735 Zeuthen, Germany}\\

\vspace{1mm}
{\it $^b$Laboratory for Theoretical Physics, JINR}\\
{\it ul.~Joliot--Curie 6, RU--141980 Dubna, Russia}\\

\vspace{1mm}
{\it $^c$Dept. of  Theoretical Physics, Faculty of Physics}\\
{\it Bishop Konstantin Preslavsky University, Shoumen, 9700, Bulgaria}\\
\end{center}
\vspace*{\fill}
\begin{abstract}
\noindent
The $O(\alpha)$ leptonic QED corrections to neutral current polarized
deep inelastic  lepton--nucleon scattering are calculated in leptonic
variables both for the case of longitudinal and transverse nucleon
polarization. The results of the complete calculation are compared with
the corresponding leading log expressions. Numerical results are
presented for the corrections in the kinematic range of the HERMES
experiment and possible future polarized proton beam experiments at HERA.

\vspace{1mm}
\noindent
PACS~: 11.10.Gh,12.15.Lk,12.15.Mm,12.20.Ds,12.20.Fv,13.60.Fz.

\end{abstract}
\vspace*{\fill}
\newpage
%
\section{Introduction}
\label{sect1}

\vspace{1mm}
\noindent
Polarized deep-inelastic lepton-hadron scattering provides one of the
cleanest methods to investigate the spin-structure of the nucleons.
This field attracted much interest  after the finding of
the EMC experiment~\cite{EMC} in 1988 that the quarks
appear to carry only a small
fraction of the nucleon spin. During the last years precise
measurements of both the polarized structure function $g_1(x,Q^2)$
of the proton and neutron by the SMC, SLAC, and HERMES
experiments~\cite{REV} were performed. Very recently also
the structure function $g_2(x,Q^2)$ was measured  for
the first time~\cite{G2}.

The unfolding of the
structure functions from the deep-inelastic scattering cross section
requires the detailed knowledge of the electromagnetic radiative
corrections, which are very large in some parts of the kinematic
domain. A first calculation of the leptonic QED corrections for the
case of pure photon exchange was performed in ref.~\cite{Minsk} for
the range of small values of $Q^2$ and a simple ansatz for the
hadronic structure functions $g_1(x)$ and $g_2(x)$~\cite{MIN1}.

In the present paper the leptonic $O(\alpha)$ QED corrections for the
neutral current deep-inelastic scattering cross sections are calculated
in a model independent approach for the case of longitudinal and
transverse nucleon polarization. We  account for both photon and
$Z$ boson exchange. The resulting corrections are therefore applicable
irrespectively of the $Q^2$ range.
This is of particular importance for  polarized
deep-inelastic scattering experiments operating in the range of
larger values of $Q^2$, e.g. possible future experiments
in the kinematic regime being accessible at HERA, cf.~\cite{JB95A}, 
but also for the
CERN fixed target experiments for the range of large $Q^2$.

The $O(\alpha)$ hadronic QED corrections are understood to be absorbed
into the parton densities, as also the  QCD
corrections~\cite{HADR1}--\cite{HADR2},
leading to $Q^2$-dependent structure functions.
In this way the hadronic tensor can be formulated quite generally
in a model independent way. Particularly the structure functions need not
necessarily  to be represented using a partonic description.

Besides the complete $O(\alpha)$ results, we also derive
the corrections in the leading logarithmic approximation (LLA).
Due to their simpler analytical structure they are particularly
suited for fast numerical estimates.

The paper is organized as follows. In section~2 the neutral current
Born cross sections are derived. The leptonic $O(\alpha)$ QED corrections
are presented in section~3. In section~4 we derive
the different contributions to the $O(\alpha)$ QED corrections in the
leading logarithmic approximation for the polarized case.
Numerical results for the kinematic
range of the HERMES experiment and possible future polarized  $ep$
collider experiments in the kinematic   regime of HERA are presented
in section~5. Section~6 contains the conclusions. An appendix
summarizes  kinematic   relations used in the calculation.
%
\section{The Born Cross Section}
\label{sect2}
\noindent
The Feynman diagram describing  neutral current deep-inelastic
lepton--nucleon scattering
\begin{equation}
l(k_1) + p(p) \rightarrow l(k_2) + X(p')
\label{eq1}
\end{equation}
is shown in Figure~1. The particle 4--momenta are given in parentheses.

The matrix element for the Born cross section reads
\ba
{\cal M}_{\rm Born}
&=& i {e^2}
         \langle p' | {\cal J}_{\mu} |p \rangle \frac{1}{Q^2}
{\bar{u}}(k_2) \left[ Q_l  \gamma^{\mu} + s_l \gamma^{\mu}
\left(v_l+a_l \gamma_5 \right) \chi(Q^2) \right] u(k_1)
\frac{1}{(2\pi)^3} \frac{1}{\sqrt{2 k^0_1  2 k^0_2}}. \nll
\label{matrixBorn}
\ea
Here $e$ denotes the electromagnetic coupling,
$q = k_1 - k_2$,
$m$ and $M$ are the lepton and nucleon masses,
${\cal J}_{\mu}$ is the hadronic current, $Q_{l}=s_l|Q_l|$
the charge of the electron,
with $s_l = 2 I^{(3)}_l$
twice the third component of the weak isospin of the electron,
$v_{l}$ and $a_{l}$ are the vector and axial-vector
couplings of the $Z$~boson to the electron
\ba
v_{l}=1-4 |Q_{l}| \sin^{2}\theta_{W}, \hspace{3.cm} a_{l}=1,
\label{veae}
\ea
and
$Q^2 = -q^2$ denotes the four momentum transfer squared.

\vspace{3cm}
\begin{picture}(120,20)(100,100) \centering
\put(135,-250){\epsfig{file=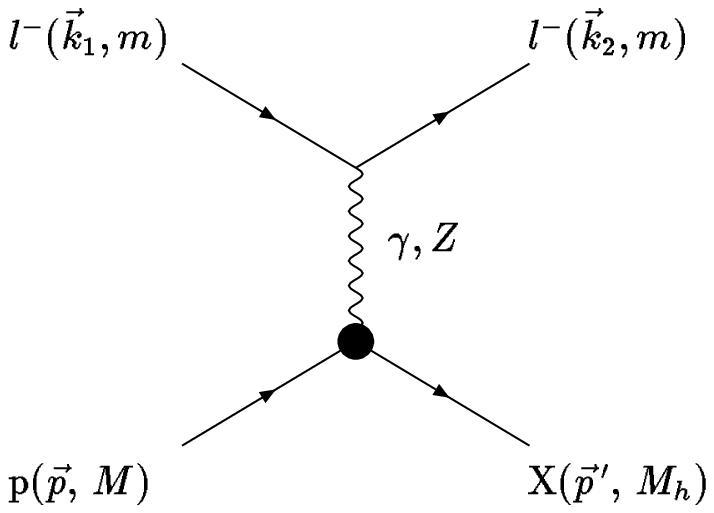,width=16cm}}
\end{picture}

\vspace{24mm}
\begin{center}
{\sf Figure~1: Born diagram for neutral current deep-inelastic
lepton--proton scattering.}
\end{center}

\vspace{2mm}
\noindent
The calculation is performed in the on--mass--shell (OMS) scheme,
in which
the weak mixing angle
$\theta_{W}$ is related to the weak boson masses by
$\sin^2 \theta_W = 1 - M_W^2/M_Z^2$.
Furthermore,
\ba
 \chi (Q^2) =
{G_\mu \over\sqrt{2}}{M_{Z}^{2} \over{8\pi\alpha}}{Q^2 \over
{Q^2+M_{Z}^{2}}}
\label{chiq}
\ea
is the $\gamma$-$Z$ propagator ratio, with
$G_{\mu}= 1.16637\cdot 10^{-5}$ GeV$^{-2}$ the Fermi
constant, and $\alpha~=~1/137.06$ the fine structure constant.

The differential Born cross section is given by:
\ba
d\sigma_{\rm Born} =
   \frac{1}{j}  (2\pi)^4  \sum_{\mr {spins}}
   \left| {\cal M}_{\rm Born} \right|^2
 {d{\vec k}_2}
\prod_i d {\vec{p}_i}\,' \delta^4 (k_1+p-k_2- \sum_i p'_i),
\label{sBorn1}
\ea
where
\ba
 p'&\equiv& \sum_i p_i' = p+k_1-k_2 \qquad  \mbox{and} \qquad
 {p'}^2 = M^2_h.
\ea
$j$ denotes the flux factor
\ba
 j= \frac{\sqrt{(k_1.p)^2-\Me\Mp}}{k^0_1 p^0 }  \frac{1}{(2\pi)^6}
  = \frac{\SLS             }{2k^0_1 p^0            (2\pi)^6},
\ea
with
\ba
\lambda_S = S^2 - 4m^2M^2,~~~~~~~~~~~~S = (p + k_1)^2 - m^2 - M^2.
\label{lambdas}
\ea
For later use we rewrite
the differential Born cross section~(\ref{sBorn1})
in terms of the
leptonic and hadronic tensors,
$L^{\mu\nu}$ and $W_{\mu\nu}$, integrating over
the phase space up to two
variables:
\ba
d\sigma_{\rm Born}
&=&  \frac {2\alpha^2}{\sqrt{\lambda_S}}\frac{1}{Q^4}
                    \Biggl[ L^{\mu\nu} W_{\mu\nu} \Biggr]
\frac{d{\vec k}_2}{k_2^0}
 =  \frac {2\pi \alpha^2}{\lambda_S}
 \frac{S^2 y}{Q^4}
 \Biggl[ L^{\mu\nu} W_{\mu\nu} \Biggr]
  dx dy.
\label{sBorn2}
\ea
$x$ and $y$ are the Bjorken variables
\be
x = \frac{Q^2}{2 p.q},~~~~~~~~~~~y = \frac{p.q}{p.k_1}.
\ee
The calculation is performed for incoming
longitudinally polarized  leptons.
We used the spin density matrix
\ba
\rho(k_1)= \sum_s u^{\small s}(k_1) {\bar u}^{\small s}(k_1) = \frac{1}{2}
           \left( 1 - \gamma_5 \hat \xi_l \right) (\hat{k_1} + m).
\ea
The 4--vector of the lepton
polarization is given by
\ba
\xi_l = \frac{\lambda_l}{m} \left (k_1 - \frac{2 m^2}{S} p \right)
\frac{S}{\sqrt{\lambda_S}},
\ea
with
\ba
\xi^2_l = - \lambda^2_l,~~~~~~~~{\rm and}~~~~~~~~~\xi_l.k_1 =0.
\ea
${\lambda_l}$ denotes the degree of lepton
polarization.
The leptonic tensor for neutral current polarized lepton  (antilepton)
scattering reads
\ba
 L^{\mu\nu} &=&   \Bigl [
     2\left(k^{\mu}_1 k^{\nu}_2+k^{\nu}_1
 k^{\mu}_2 \right )
- g^{\mu \nu} Q^2  \Bigr ]
   L_{S} (Q^2,\lambda_l)
 -2 i  k_{ 1 \alpha} k_{ 2 \beta} \varepsilon^{\alpha\beta\mu\nu}
   L_{A} (Q^2,\lambda_l) \nonumber\\
& &  + 4   \frac{m^2}{S} \Biggl \{
i~\varepsilon^{\alpha \beta \mu \nu}
\Biggl  [
p_{\alpha} q_{\beta}
L_v(Q^2,\lambda_l)
-p_{\alpha} \left (k_{1\beta} + k_{2\beta}\right)
 L_a(Q^2,\lambda_l)  \Biggr ]
\nonumber\\
& &     +   \Biggl  [ p^{\mu} \left ( k_{1}^{\nu} +  k_{2}^{\nu} \right )
          + p^{\nu} \left ( k_{1}^{\mu} +  k_{2}^{\mu} \right )
          - p^{\mu} q^{\nu}
          - p^{\nu} q^{\mu}  \nonumber\\
& &     - g^{\mu\nu} \Bigl [ p.(k_1 +
           k_2) - p.q \Bigr ] \Biggr ]
\Biggr \}
L_{\chi}(Q^2,\lambda_l)
+ 4 g^{\mu\nu} m^2
\Bigl [L_{\chi}(Q^2,\lambda_l) - L_a(Q^2,\lambda_l)\Bigr ]~.
\ea
The symmetric $(S)$ and  antisymmetric $(A)$ parts are
\ba
   L_{S} (Q^2,\lambda_l)  =          Q^2_{l} + 2|Q_l|
               \left( v_l - \Pe \lambda_l a_l  \right) \chi(Q^2)
      + \left( v^2_l + a^2_l
- 2 \Pe  \lambda_l v_l a_l \right) \chi^2(Q^2),
\ea
\ba
   L_{A} (Q^2,\lambda_l)=
       - \lambda_l
        Q^2_l + 2 |Q_l|
        \left( p_l a_l - \lambda_l v_l \right) \chi(Q^2)
 +\Biggl(2 p_l v_l a_l -  \lambda_l \left( v^2_l + a^2_l\right)
             \Biggr) \chi^2(Q^2)~.
\label{lnrten}
\ea
The formfactors contributing to the terms
$\propto m^2$ are
\ba
L_v(Q^2,\lambda_l) &=& \lambda_l
\left[|Q_l|+v_l\chi(Q^2)\right]^2,
\nll
L_a(Q^2,\lambda_l) &=& \lambda_l
a^2_l\chi^2(Q^2),
\nll
L_{\chi}(Q^2,\lambda_l) &=& \lambda_l p_l
a_l \chi(Q^2)\left[|Q_l| + v_l \chi(Q^2)\right].
\ea
The particle label  $\Pe$ takes the values
$\Pe = 1$
for particles  and
$-1$ for antiparticles.

The hadronic tensor reads
\ba
 W_{\mu\nu}     &=& p^0 (2\pi)^6 \sum   \int
          \langle p'| {\cal J}_{\mu} |p     \rangle
          \langle p    | {\cal J}_{\nu} |p' \rangle
          \delta^4 (p' - \sum_i p'_i)
                            \prod_i d{\vec{p_i}\,'}.
\ea
For its representation in terms of structure functions we follow the
convention of ref.~\cite{BK},
\ba
W_{\mu\nu}  &=& \left(-g_{\mu\nu}+ \frac{q_\mu q_\nu }{q^2} \right)
                                              {\cal F }_1(x,Q^2)
 +\frac{\widehat{p_\mu}\widehat{p_\nu}}{p.q}
                                              {\cal F }_2(x,Q^2)
     - i \varepsilon_{\mu\nu\lambda\sigma} \frac{q^\lambda p^\sigma}
                                { 2 p.q}  {\cal F }_3(x,Q^2)       \nll
  && + i \varepsilon_{\mu\nu\lambda\sigma}
\frac{q^\lambda {s}^\sigma} {p.q}{\cal G }_1(x,Q^2)
     + i \varepsilon_{\mu\nu\lambda\sigma}
          \frac {q^{\lambda}(p.q {s}^\sigma
                      -{s}.q p^\sigma)}{(p.q)^2}
                                            {\cal G }_2(x,Q^2)      \nll
  && +      \left[\frac{\widehat{p_\mu} \widehat{{s}_\nu}
        +   \widehat {{s}_\mu} \widehat{p_\nu}}{2}
                -  {s}.q
            \frac{\widehat{p_\mu}\widehat{ p_\nu}}{p.q}\right]
                     \frac{1}{p.q}                {\cal G }_3(x,Q^2)\nll
  && +   {s}.q \frac{\widehat{p_\mu} \widehat{p_\nu}}{(p.q)^2}
                                                  {\cal G }_4(x,Q^2)
               +  \left( - g_{\mu\nu}+ \frac{q_\mu q_\nu }{q^2}\right)
                  \frac{ {s}.q  }{p.q}       {\cal G }_5(x,Q^2),
\label{hadten}
\ea
where
\ba
 \widehat {p_\mu}   =   p_\mu - \frac{p.q }{q^2}  q_\mu,\qquad
 \widehat {{s}_\mu}  =  {s}_\mu - \frac{s.q}{q^2}  q_\mu.
\ea
$s$  denotes the polarization 4--vector of the nucleon.
In the nucleon rest frame it is given by
\ba
    s= M(0,{\vec{n}}_{\lambda}).
\ea
${\vec n}_\lambda $
is an unit $3$-vector.

For a short-hand notation we have introduced the
combined neutral current structure functions ${\cal F}_i$ and ${\cal G}_
i$
in eq.~(\ref{hadten}).
In terms of the structure functions  $F^{{J_1 J_2}}_i$ and
$g^{{J_1 J_2}}_i$,
which are associated with the respective currents, they read:

\ba
{\cal  F }_{1,2}(x,Q^2)
 &=& Q^2_{l} F^{\gamma\gamma}_{1,2}(x,Q^2)
      + 2 |\Ql|
 \left( v_l -\pe \lambda_l a_l \right)\chi(Q^2) F^{\gamma Z}_{1,2}(x,Q^2
) \nll\nll
      &&+
 \left( v^2_l + a^2_l - 2\pe\lambda_l v_l a_l \right) \chi^2(Q^2)
                                                  F^{ZZ}_{1,2}(x,Q^2),
  \\  \nll
{\cal F }_{3}(x,Q^2)   &=&
 2|Q_l|\left( p_l a_l - \lambda_l v_l \right) \chi(Q^2)
                                                  F^{\gamma Z}_3(x,Q^2)
  \nll
      &&+ \left[2 p_l v_l a_l -  \lambda_l
          \left( v^2_l + a^2_l\right)\right] \chi^2(Q^2)
                                                  F^{ZZ}_3(x,Q^2) ,
  \\  \nll
{\cal G }_{1,2}(x,Q^2) &=&
 -Q^2_{l}\pe \lambda_l  g^{\gamma \gamma}_{1,2}(x,Q^2)
     + 2 |Q_l| \left( p_l a_l - \lambda_l v_l \right)\chi(Q^2)
     g^{\gamma Z}_{1,2}(x,Q^2)  \nll
  && +  \left[2 p_l v_l a_l - \lambda_l
         \left( v^2_l + a^2_l \right)\right] \chi^2(Q^2)
    g^{ZZ}_{1,2}(x,Q^2) , \\  \nll
{\cal G }_{3,4,5}(x,Q^2) &=&
              2 |Q_l|\left( v_l -\pe \lambda_l a_l \right) \chi(Q^2)
         g^{\gamma Z}_{3,4,5}(x,Q^2)   \nll\nll
  && +  \left( v^2_l + a^2_l - 2\pe\lambda_l v_l a_l \right)
                                         \chi^2(Q^2)
                                        g^{ZZ}_{3,4,5}(x,Q^2).
\ea
The twist-$2$ contributions to the structure functions can be expressed
in terms of parton densities.
In lowest order QCD one obtains~\cite{R1,R3,BK}~:
\ba
 F^{{J_1 J_2}}_1(x,Q^2) =
 \sum_q \alpha^q_{{J_1 J_2}} \left[ q(x,Q^2)+\bar q(x,Q^2) \right],
\ea
\ba
 F^{{J_1 J_2}}_2(x,Q^2) = {2x}{F^{{J_1 J_2}}_1(x,Q^2)},
\ea
\ba
 F^{{J_1 J_2}}_3(x,Q^2) =
 \sum_q \beta^q_{{J_1 J_2}} \left[ q(x,Q^2)-\bar q(x,Q^2) \right],
\ea
\ba
\label{eqg1}
 g^{{J_1 J_2}}_1(x,Q^2) &=&
\frac{1}{2}
 \sum_q \alpha^q_{{J_1 J_2}}
\left[\Delta q(x,Q^2) + \Delta \bar q(x,Q^2) \right],
\\ \nll
 g^{{J_1 J_2}}_2(x,Q^2) &=& -g^{{J_1 J_2}}_1(x,Q^2)
                       +\int_x^1\frac{dy}{y} g^{{J_1 J_2}}_1(y,Q^2),
\\ \nll
 g^{{J_1 J_2}}_3(x,Q^2) &=& 4x
                          \int_x^1\frac{dy}{y} g^{{J_1 J_2}}_5(y,Q^2),
\\ \nll
 g^{{J_1 J_2}}_4(x,Q^2) &=& 2 x g^{{J_1 J_2}}_5(x,Q^2),
\\ \nll
\label{eqg5}
 g^{{J_1 J_2}}_5(x,Q^2) &=&
 \sum_q \beta^q_{{J_1 J_2}}
\left[\Delta q(x,Q^2)-\Delta\bar q(x,Q^2) \right],
\ea
where
\ba
\alpha^q_{{J_1 J_2}} &=&
\alpha^q_{{\gamma\gamma,\gamma Z ,ZZ}}
=\left[e^2_q,~~2e_q v_q,~~v_q^2+a_q^2\right], \\
 \beta^q_{{J_1 J_2}} &=&
\beta^q_{{\gamma\gamma,\gamma Z ,ZZ}}
=\left[0 ,~~2e_q a_q,~~2v_q a_q \right],
\ea
and the electroweak couplings are
\renewcommand{\arraystretch}{2}
\begin{equation}
\begin{array}{ll}
 e_u  = + {\ds \frac{ 2}{ \ds 3} },  &
 e_d  =
 - {\ds
\frac{    1}{    3} },  \\
 v_u  =  {\ds
  \frac{    1}{    2}}
- {\ds
\frac{    4}{    3}} \sin^2\theta_W,   &
v_d  =  - {\ds
\frac{    1}{    2}}
+ {\ds
\frac{    2}{    3}}
\sin^2\theta_W,   \\
 a_u  = {\ds
  \frac{    1}{    2} }, &
 a_d  =  - {\ds
\frac{    1}{    2}}.
\end{array}
\end{equation}
For the numerical results, which are presented in section~5 below,
the structure functions~eq.~(\ref{eqg1}--\ref{eqg5}) are parametrized
using a partonic description for the structure functions $F_1(x,Q^2),
F_3(x,Q^2), g_1(x,Q^2)$
and $g_5(x,Q^2)$. The other structure functions are calculated using
the relations above.
$q(x,Q^2)$, $\overline{q}(x,Q^2)$, $\Delta q(x,Q^2)$, and
$\Delta \overline{q}(x,Q^2)$ denote the unpolarized and
polarized quark and antiquark densities, respectively.

It appears to be convenient to rewrite
the Born cross section in terms of the following two
contributions:
\ba
\frac{ d^2\sigma_{{\mr Born}}}{ dx dy}  =
\frac{ d^2\sigma^{{\mr unpol}}_{{\mr Born}}}{ dx dy}
+\frac{ d^2\sigma^{{\mr  pol}}_{{\mr Born}}}{ dx dy},
\ea
with
\ba
\frac{ d^2\sigma^{{\mr unpol}}_{{\mr Born}}}{ dx dy}
 =    \frac{2\pi \alpha^2 }{Q^4} S
\sum_{i=1}^{3} \;  S_i^U(x,y) {\cal F}_i(x,Q^2),
\label{equpol}
\ea
and
\ba
\frac{ d^2\sigma^{{\mr   pol}}_{{\mr Born}}}{ dx dy}
 =    \frac{2\pi \alpha^2 }{Q^4}
     \lambda^p_N f^{p} S
     \sum_{i=1}^{5} \;  S^p_{gi}(x,y) {\cal G}_i(x,Q^2).
\label{eqpol}
\ea
$\lambda_N^p$ denotes the degree of nucleon polarization.
For unpolarized deep-inelastic scattering only the first term,
$d^2 \sigma^{\mr unpol}$, contributes.
Eq.~(\ref{eqpol}) applies both to the case of
longitudinal $(L)$ and
transversal $(T)$ nucleon polarization, where
\ba
f^L &=& 1,                                            \\
f^T &=&               \COSPHI \;\frac{ d\varphi}{2\pi}
\sqrt{\frac{4\Mp x}{Sy}\left(1-y-\frac{\Mp x y}{S}\right)}
\equiv \COSPHI \;\frac{ d\varphi}{2\pi}
\frac{1-y}{y}\sin\theta_2.
\label{Borntran}
\ea
$\theta_2$ is the angle between the incoming and outgoing leptons.
$\varphi$ denotes
 the angle between the nucleon spin vector $s$ and
the plane of the incoming  and outgoing
lepton  in the nucleon rest frame (see  figure~4 in appendix~A).
The polarization 3--vectors for longitudinal and transverse polarization
are given by
\ba
\vec{n}^{{L }} &=& \lambda_N^L\frac{\vec{k}_1}{|\vec{k}_1|},\\
\vec{n}^{{T }} &=&
\lambda_N^T
\vec{n}_{\bot }, \quad \mbox{with}  \quad
\vec{n}_{\bot }
\vec{k}_{1    }=0.
\ea
Finally,
the kinematic   coefficients in eqs.~(\ref{equpol}) and (\ref{eqpol})
are:
\ba
  S_1^U(x,y) &=&  2 x y^2,                              \nll
  S_2^U(x,y) &=&   2 \left[ (1-y)-xy \frac{\Mp}{S}\right],    \nll
  S_3^U(x,y) &=&   x [1 - (1 - y)^2],
\ea
\ba
   S^{L}_{1}(x,y) &=&
                          2xy
        \Biggl[ ( 2 - y )-2 \frac{\Mp}{S} xy \Biggr]  ,
                                                             \nll
   S^{L}_{2}(x,y) &=&
                             -8 x^2 y \frac{\Mp}{S}  ,
                                                             \nll
   S^{L}_{3}(x,y) &=&
                   4\frac{\Mp}{S}
        \Biggl[(1 - y) - xy \frac{\Mp}{S} \Biggr],
                                                             \nll
   S^{L}_{4}(x,y) &=&
                        -2
        \left(1+\frac{2x\Mp}{S}\right)
        \Biggl[  \left( 1 - y \right)-xy \frac{\Mp}{S} \Biggr],
                                                             \nll
   S^{L}_{5}(x,y)&=&  -2xy \left(  y + 2 \frac{\Mp}{S} x y \right),
\ea
\ba
   S^{T}_{1}(y,Q^2) &=&    S^{T}_{5}(y,Q^2)  =
   S_1^U(y,Q^2)  ,                                 \nll
   S^{T}_{2}(y,Q^2) &=& 4 x y,                         \nll
   S^{T}_{3}(y,Q^2) &=&-\frac{S^{L}_{1}(y,Q^2)}{2xy}, \nll
   S^{T}_{4}(y,Q^2) &=& S_2^U(y,Q^2).
 \ea
%
\section{The $O(\alpha)$ Leptonic Correction}
\label{sect3}
%

\vspace{1mm}
\noindent
The $O(\alpha)$ leptonic radiative corrections to the scattering cross
sections are given by
\be
\frac{d^2 \sigma^{\rm QED, 1}_{\rm rad}}{d x_l d  y_l} =
\frac{\alpha}{\pi} \delta_{\rm VR}
\frac{d^2 \sigma_{\rm Born}}{d x_l d y_l} +
\frac{d^2 \sigma_{\rm Brems}}{d x_l dy_l} =
\frac{d^2 \sigma_{\rm rad}^{\rm unpol}}{d x_l d y_l} +
\frac{d^2 \sigma_{\rm rad}^{\rm pol}}{d x_l dy_l}.
\label{COR1}
\ee
The corrections are represented by the finite parts of the
virtual and soft (VR) and Bremsstrahlung terms, respectively.
Since we integrate over the phase space of the radiated photon, the
differential cross sections~(\ref{COR1}) depend on the way in which
the Bjorken variables $x$ and $y$ are determined kinematically.
Because in all the polarized deep-inelastic scattering experiments
performed so far the kinematic   variables were measured using the
scattered lepton\footnote{Other choices of kinematic   variables were
also
investigated for the case of unpolarized deep-inelastic
scattering~\cite{JB1,JB2,R1,R3}.} the present calculation refers
to this set of variables. The kinematic variables are defined by
\be
x_l = \frac{Q^2_l}{S y_l},~~~~y_l =
\frac{p_1.(k_1 - k_2)}{p_1.k_1},~~~~{\rm and}~~Q_l^2 = (k_1 - k_2)^2.
\label{eqKIN1}
\ee
The hadronic
structure functions depend  on the hadronic variables
\be
x_h = \frac{Q^2_h}{S y_h},~~~~y_h =
\frac{p_1.(p_2 - p_1)}{p_1.k_1},~~~~{\rm and}~~Q_h^2 = (p_2 - p_1)^2,
\label{eqKIN2}
\ee
over which is integrated.

\subsection{Virtual and Soft Corrections}
\label{sect31}

\vspace{3cm}
\begin{picture}(120,20)(100,100) \centering
\put(135,-250){\epsfig{file=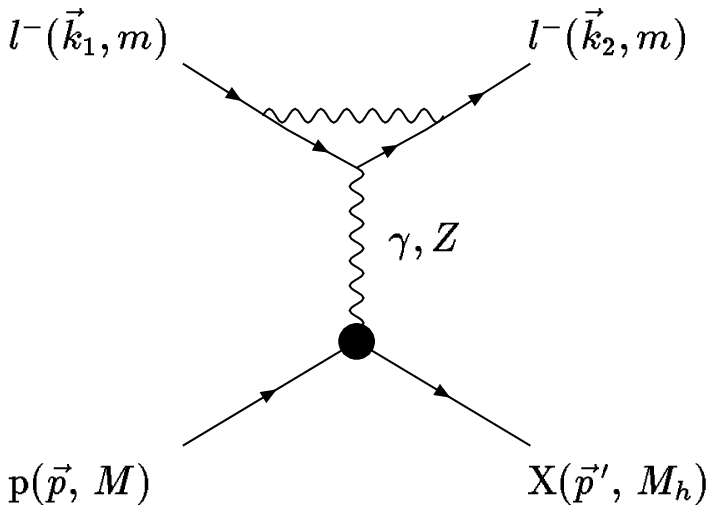,width=16cm}}
\end{picture}

\vspace{24mm}
\begin{center}
{\sf Figure~2: Diagram of the leptonic
virtual photon correction for deep-inelastic
scattering.}
\end{center}

\vspace{2mm}
\noindent
The virtual and soft correction, $\delta_{\rm VR}$, can be
written as
\ba
\delta_{\rm VR}(y_l, Q^2_l)
 &=&  \delta_{\rm inf}(y_l, Q^2_l) - \frac{1}{2} \ln^2 \left [
\frac{1 - y_l(1 - x_l)}{1 - y_l x_l} \right ]
+ {\rm Li}_2 \left [ \frac{1 - y_l}{(1 - y_l x_l)(1 - y_l(1 -x_l))}
\right ] \nonumber\\ & &~~~~~~+ \frac{3}{2} \ln \left ( \frac{Q^2_l}{m^2} \right ) -
{\rm Li}_2(1)  - 2,
\ea
with
\be
\delta_{\rm inf}(y_l, Q^2_l) =  \left [
\ln \left(
\frac{Q^2_l}{m^2}
\right ) - 1 \right ]
\ln \left [
\frac{y_l^2 (1 - x_l)^2}{(1 - y_l x_l)[1 - y_l(1 - x_l)]}
\right ]
\ee
and
\be
{\rm Li}_2(x) = - \int_0^1 dz \frac{\ln(1 - xz)}{z}.
\ee
These expressions are the same in the unpolarized and polarized case.
They were derived in refs.~\cite{R1,R3}.
%
\subsection{Bremsstrahlung Corrections}
\label{sect32}

\vspace{1mm}
\noindent
The differential Bremsstrahlung
cross section for the scattering of polarized electrons
off polarized protons, originating from the  diagrams  in figure~3,
is
\ba
\frac{\ds d\sigma_{{\mr Brem}}}{\ds d\xl d\yl}
=  { 2  \alpha^3 S\yl}
\int d\yh d\Qh \frac{1}{\QhS}
\Biggl[\frac{1}{2\pi}\frac{d\varphi_k}{{\SLQ}}
\frac{1}{4}\Biggl(L^{\mu\nu}_{{\rm rad}} W_{\mu\nu} \Biggr)\Biggr]
\label{cr_sect_rad}.
\ea

\vspace{3cm}
\begin{picture}(120,20)(100,100) \centering
\put(135,-250){\epsfig{file=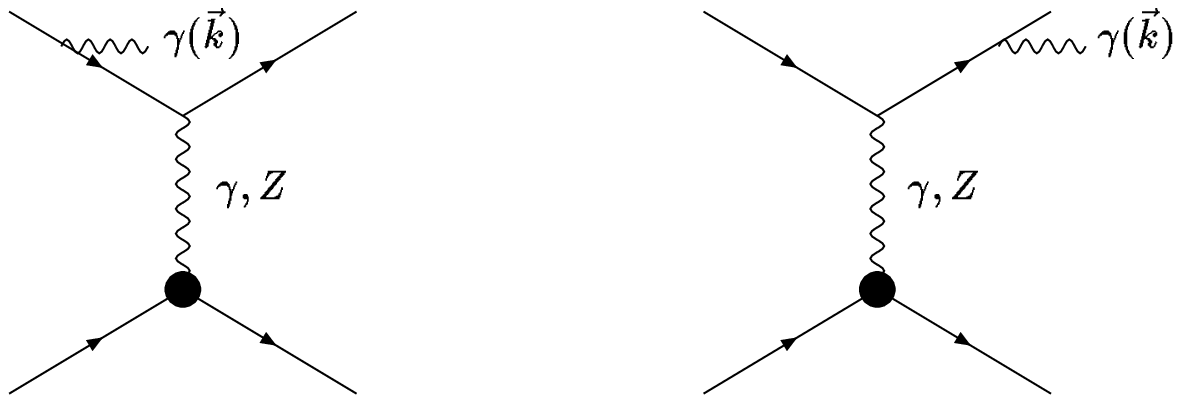,width=16cm}}
\end{picture}

\vspace{24mm}
\begin{center}
{\sf Figure~3: The leptonic Bremsstrahlung diagrams for
for deep-inelastic scattering.}
\end{center}

\vspace{2mm}
\noindent
$W_{\mu \nu}$ denotes the hadronic tensor~(\ref{hadten}). The invariant
$\lambda_q$ is defined in (\ref{eqLQ}).
The Bremsstrahlung correction to the leptonic tensor,
$L^{\mu\nu}_{\rm{rad}}$ has the following form~:
\ba
L^{\mu\nu}_{\rm{rad}} &=&
        2 L_{_S}(Q^2_h)
\Biggl\{ 4 \Biggl( -\frac{2\Me}{z^2_2}
+\frac{\Qh}{z_1 z_2}
\Biggr) k_{1}^{\mu} k_{1}^{\nu}
       + 4 \Biggl( -\frac{2\Me}{z^2_1}
+\frac{\Qh}{z_1 z_2}
\Biggr) k_{2}^{\mu} k_{2}^{\nu}
\nll
&&
 +   \Biggl [ 2 \Me \Qh \Biggl( \frac{1}{z^2_1}
+ \frac{1}{z^2_2} \Biggl) - \frac{Q_l^4 + Q_h^4}{z_1 z_2}
- 2  \Biggr] g_{\mu\nu}
\Biggr\}
\nll
&&       +  4 i
\varepsilon^{\alpha\beta\mu\nu}  \Biggl \{ - L_A(Q^2_h,\lambda_l)
\Biggl [
  \Biggl( \frac{2\Me}{z^2_2}-\frac{\Ql}{z_1 z_2}
-\frac{1}{z_2} \Biggr)
k_{1\alpha} q_{ \beta}
 + \Biggl( \frac{2\Me}{z^2_1}-\frac{\Ql}{z_1 z_2}
+\frac{1}{z_1} \Biggl)
k_{2\alpha} q_{ \beta}
\Biggr ]
\nll
&& + 2 \left [ L_v(Q^2_h,\lambda_l) + L_a(Q^2_h,\lambda_l)\right]
\frac{\Me}{z^2_1}
\Biggl [
      \ylh
k_{1\alpha} k_{2\beta}
 +   \left(  1 +  \ylh \right)
k_{2\alpha} k_{ \beta}
\Biggr ]
\nll
&&
       -  2   \left[ L_v(Q^2_h,\lambda_l) - L_a(Q^2_h,\lambda_l)\right]
\frac{\Me}{z^2_1}
k_{1\alpha} k_{ \beta}  \Biggr \}
\nll
&&
       +   8 L_{\chi}(Q^2_h,\lambda_l) \frac{\Me}{z^2_1}
\Biggl [ 2 \left( k_{1}^{\mu} k_{2}^{\nu}+k_{1}^{\nu} k_{2}^{\mu} \right)
       - 4 \left( 1 + \ylh \right)  k_{2}^{\mu} k_{2}^{\nu}
       -   \left( \Qlh - \ylh \Qh \right) g^{\mu\nu}
\Biggr ]~.
\nonumber
\label{lrtaft}
\ea
Here we retained only those terms in $O(m^2)$, which yield non-vanishing
contributions after intgrating over $\varphi_k$ in the limit
$m \rightarrow 0$.
Similar to the leptonic tensor in the Born approximation,
the tensor $L^{\mu \nu}_{\rm    rad}$ can be decomposed into a symmetric
and an asymmetric term, and also
a  contribution $\propto m^2$.
Besides the invariants defined in
eqs.~(\ref{eqKIN1},\ref{eqKIN2}), it depends on
\be
z_1 = 2 k.k_1,~~~~~~~~~~z_2 = 2 k.k_2.
\label{z12}
\ee
The finite part of the Bremsstrahlung contributions are  expressed
in form of integrals over the radiator functions
$S_X^i(y_l, Q_l^2, y_h, Q^2_h)$
and
${\cal S}_Y^i(y_l, Q_l^2, y_h, Q^2_h)$.
Since the structure of the differential cross sections turns out to be
partly different
for  unpolarized and polarized deep-inelastic scattering,
we will describe the individual cases separately.
\subsubsection{Unpolarized Case}
\label{sect321}

\vspace{1mm}
\noindent
The finite contribution to the Bremsstrahlung cross section
reads
\ba
\frac{ d\sigma_{{\mr Brem}}}{ d\xl d\yl}
&=&
  2  \alpha^3 S \int_{\bf R} d\yh d\Qh
\Biggl \{
\frac{1}{\QhS}
\sum_{i=1}^{3} \;
  {\cal S}_i^U(\yl,\Ql,\yh,\Qh) {\cal F}_i(\xh,\Qh)
\nonumber\\
& &~~~~~-\frac{1}{Q_l^4} \sum_{i=1}^3
  S_i^U(\yl,\Ql)
{\cal L}^{\rm IR}(\yl,\Ql,\yh,\Qh)
{\cal F}_i(\xl,\Ql) \Biggr \}~,
\label{eqBREM}
\ea
where
\ba
{\cal L}^{\mr {IR}}(\yl,\Ql,\yh,\Qh)
&=&
\frac{\Ql+2m^2}{\Qlh} \Biggl(\ \frac{1}{\sqrt{C_1}}
   -\frac{1}{\sqrt{C_2}} \Biggr)\
   -m^{2}\Biggl(\ \frac{B_{1}}{C_{1}^{3/2}}
                +\frac{B_{2}}{C_{2}^{3/2}} \Biggr).
\label{eqn052a}
\ea
The integral over $\varphi_k$ in eq.~(\ref{cr_sect_rad})
can be performed analytically, see appendix~A. The remaining twofold
integral takes the form~\cite{R1}
\ba
\int_{\bf R} d y_h d Q^2_h = \sum_{l=1}^2
\int_{\hat{y}_{h_l}}^{\hat{y}_{h_{l+1}}} d y_h
\int_{Q^2_{h_{1,d}}}^{Q^2_{h_{l,u}}} d Q^2_h,
\ea
with
\ba
\hat{y}_{h_1} &=& \frac{y_l(y_l S - \sqrt{\lambda_q}) + 2 Q^2_l M^2/S}
               {    y_l S - \sqrt{\lambda_q}  +         M^2  } \nll
\hat{y}_{h_2} &=& \frac{y_l(y_l S + \sqrt{\lambda_q}) + 2 Q^2_l M^2/S}
               {    y_l S + \sqrt{\lambda_q}  +         M^2  } \nll
\hat{y}_{h_3} &=& y_l \nll
Q^2_{h_{1,d}} &=& Q^2_l + \frac{S}{2 M^2} y_{lh} (y_l S -
\sqrt{\lambda_q}) \nll
Q^2_{h_{2,u}} &=& Q^2_l + \frac{S}{2 M^2} y_{lh} (y_l S +
\sqrt{\lambda_q}) \nll
Q^2_{h_{1,u}} &=& y_h S,
\ea
and
\ba
 \ylh  &=& \yl-\yh.
\ea
The functions   $S_i^U(\yl,\Ql,\yh,\Qh)$ are given by~:
\ba
 S_1^U(\yl,\Ql,\yh,\Qh) &=& \yl \Biggl\{
       - 2\Me \Qh \left( \BCl+\BCll \right) \nll
       && + \frac{\QlS+\QhS}{\Qlh}
         \left(\lClS - \lCllS  \right)
       + \frac{2}{\SLQ} \Biggr\},
\\
 S_2^U(\yl,\Ql,\yh,\Qh)
 &=& \frac{\yl}{S\yh}
    \Biggl\{
- 2\Me\Biggl[
     \Biggl(S^2\yll (\yll+\yh)-\Mp \Qh\Biggr)   \BCl                  \nll
    && +\Biggl(S^2 \yhl-\Mp \Qh\Biggr)       \BCll \Biggr]            \nll
    && +\Biggl[S^2\Qh (\ylpl-\yl\yh)-\Mp (\QlS+\QhS)\Biggr]
        \frac{1}{\Qlh}  \left(\lClS - \lCllS  \right)              \nll
    && -S^2\yh\Biggl[ \lClS
                 + \frac{\yll}{\CllS} \Biggr]
       - \frac{2 \Mp }{\SLQ}
    \Biggr\},
\\
S_3^U(\yl,\Ql,\yh,\Qh) &=&
 \frac{\yl}{\yh} \Biggl\{
  - \Me  \Qh
      \Biggl[ (2 \yll+\yh)\BCl+(2-\yh)\BCll \Biggr]                   \nll
&&+ (2-\yl) \frac{\Ql\Qh}{\Qlh}\left(\lClS - \lCllS  \right)      \nll
&&- \Qh                   \left(\frac{\yll}{\ClS}  - \lCllS \right)
  - \frac{\yh(\Ql+\Qh)}{2}\left(\lClS              + \lCllS\right)
                              \Biggr\}~,
\ea
with
\ba
 \yll  &=& 1-\yl,          \nll
 \yhl  &=& 1-\yh, \nll
 Q^2_{lh} &=& Q^2_l - Q^2_h.
\ea
The coefficients $B_{1,2}$ and $C_{1,2}$ are defined in
eqs.~(\ref{eqB1}--\ref{eqC2}).
These functions were given in refs.~\cite{R1,R3} before and are
presented here in a somewhat more compact form.

\subsubsection{Longitudinal Polarization}
\label{sect322}

\vspace{1mm}
\noindent
The Bremsstrahlung part of the differential cross section takes the form
\ba
\frac{\ds d\sigma^{{\mr {Long}}}_{{\mr {Brem}}}}{\ds d\xl d\yl}
&=&
  2 \alpha^3 \lambda_N^L \int d\yh d\Qh \Biggl\{
\frac{1}{\QhS}
\sum_{i=1}^{5}\;S^{L}_{i}(\yl,\Ql,\yh,\Qh) {\cal G}_i(\xh,\Qh) \nll
& &~~~~-\frac{1}{Q^4_l}
\sum_{i=1}^{5}\;S^{L}_{i}(\yl,\Ql)
{\cal L}^{\rm IR}(y_l,Q^2_l,y_h,Q^2_h)
{\cal G}_i(\xh,\Qh) \nll
& &~~~~+ 2  \frac{\Me}{Q_h^4} \BCl \Biggl[
 \sum_{i=1}^{2}\Biggl(
   {\cal S}^{L}_{vi}(\yl,\Ql,\yh,\Qh){\cal G}^v_i(\xh,\Qh)
  +{\cal S}^{L}_{ai}(\yl,\Ql,\yh,\Qh){\cal G}^a_i(\xh,\Qh)\Biggr)
\nll
& &~~~~+\sum_{i=3}^{5}
   {\cal S}^{L}_{\chi i}(\yl,\Ql,\yh,\Qh){\cal G}^{\chi}_i(\xh,\Qh)
\Biggr]
\Biggr\}~.
\ea
As in the unpolarized case, to
each of the structure functions corresponds   a kinematic function
$S^{L}_{i}(\yl,\Ql,\yh, \Qh)$.
Furthermore, infrared-finite terms
$\propto m^2$  contribute, which are denoted by
${\cal S}^{L}_{vi,ai}(\yl,\Ql,\yh, \Qh)$, and
${\cal S}^{L}_{\chi i}(\yl,\Ql,\yh, \Qh)$, respectively.

These coefficients are given by~:
\ba
S^{L}_{1}(\yl,\Ql,\yh, \Qh) &=&
     \frac{\yl}{S\yh} \Biggl\{
  \left[ S^2(2\yll+\yh)-2 \Mp \Ql\right] \Qh
        \nll
                  && \times
  \left[-2 \Me \BCl
 + \frac{\Ql}{\Qlh}\left(\lClS -\lCllS \right)  \right]
        \nll
  && + \left[ S^2 (2-\yh)- 2 \Mp \Qh\right]\Qh
        \nll
  && \times  \left[- 2 \Me \BCll
   + \frac{\Ql}{\Qlh} \left(\lClS -\lCllS \right) \right]
        \nll
  && - \left[ S^2\left( \yh \Ql
     +(2\yll+\yh)\Qh \right) + 2 \Mp \QlS \right]
         \frac{1}{ \ClS }
        \nll
  && - \left[ S^2\left(\yh\Ql+(-2+\yh)\Qh\right)
 + 2\Mp \QhS\right]
         \frac{1}{ \CllS}
        \nll
  && - \frac{4 \Mp ( \Ql + \Qh )}{ \SLQ}
                          \Biggr\},
\ea
\ba
 S^{L}_{2}(\yl,\Ql,\yh, \Qh) &=&
   4 \frac{\yl}{S\yhS}\Mp \Qh \Biggl\{
      (\yll+\yh) \Ql
    \Biggl[ 2\Me\BCl+\lClS
            \nll
    &&  -\frac{\Ql}{\Qlh}
    \left(\lClS -\lCllS \right) \Biggr]
            \nll
    && + \Qh \left[
          2 \Me \BCll - \lCllS
 - \frac{\Ql}{\Qlh}\left(\lClS -\lCllS \right) \right]
         \nll
    && - \Ql          \left(\lClS +\frac{\yll}{\CllS}\right)
       - \frac{\yl}{\SLQ}
                                           \Biggr\},
\ea
\ba
 S^{L}_{3}(\yl,\Ql,\yh, \Qh) &=&
     \frac{\yl}{S^2\yhS} {\Mp} \Biggl\{
        \Biggl[S^2 \Biggl(\yh \Qh-(\yll+\yh) (2\yll+\yh)\Ql\
\Biggr)
                                + 2\Mp \Ql \Qh\Biggr]
             \nll
    &&\times  \Biggl[ 2\Me \BCl
             - \frac{\Qh}{\Qlh} \left(\lClS - \lCllS \right)
\Biggr]   \nll
    &&   - 2\Qh \left( S^2\yhl-\Mp \Qh \right)
             \nll
    &&\times  \Biggl[ 2\Me  \BCll
             - \frac{\Qh}{\Qlh} \left(\lClS - \lCllS \right)
\Biggr]   \nll
    &&   +     { 2 \left[S^2 \left(-\yh \Ql+ \yhl\Qh\right)
                            -\Mp (\QlS+\Ql\Qh+\QhS)\right]}
               {\lClS}
             \nll
    &&   +     {\left[S^2\left[\yh\yhl \Ql+\left(2\yllS
                +\yh\right)\Qh\right] +2\Mp\Ql\Qh\right]}
              {\lCllS}
             \nll
    &&   +\left[S^2 (-4+\yl)\yh
         -4\Mp \left( \Qh+\frac{B_2}{\SLQS} \right) \right]
 \frac{1}{\SLQ}
                                         \Biggr\},
\ea
\ba
 S^{L}_{4}(\yl,\Ql,\yh, \Qh)
 &=& \frac{\yl}{S^2\yhS} \Biggl\{
       \left(S^2\yh+2\Mp \Ql\right)
       \left[S^2\yll (\yll+\yh)-\Mp\Qh\right]
             \nll
 && \times \Biggl[ 2\Me \BCl
  - \frac{\Qh}{\Qlh} \left(\lClS - \lCllS \right) \Biggr]
          \nll
 &&  + \left(S^2\yh+2\Mp\Qh\right)
         \left[ S^2 \yhl-\Mp \Qh \right]
             \nll
 && \times  \Biggl[ 2\Me \BCll
       - \frac{\Qh}{\Qlh} \left(\lClS -\lCllS\right)   \Biggr]
    \nll
 &&  + \Biggl[\left(S^2\yh+2\Mp \Ql\right) \left[S^2\yh
     + \Mp (\Ql+\Qh)\right]
             \nll
 &&     -2 \Mp \Qh \left[S^2 \yhl - \Mp \Qh \right]\Biggr] \lClS
       \nll
 &&  + \Biggl[\left( S^2\yh+2\Mp \Qh\right)
             \left[S^2\yh \yll-\Mp (\Ql+\Qh) \right]
             \nll
 &&  - 2\Mp \Qh \left[S^2\yll (\yll+\yh)-\Mp\Qh\right]\Biggr
]\lCllS      \nll
 &&  + 2\Mp\left[ S^2 (2+\yll)\yh
     + 2\Mp\left(\Qh+\frac{B_2}{\SLQS}\right) \right]\frac{1
}{\SLQ}
                                         \Biggr\},
\ea
\ba
 S^{L}_{5}(\yl,\Ql,\yh, \Qh)
 &=& \frac{\yl}{S \yh} \Biggl\{
       ( S^2 \yh  + 2\Mp \Ql)
             \nll
 && \times \Biggl[ 2 \Me \Qh \BCl
 -  \frac{\Ql \Qh}{\Qlh}\left(\lClS -\lCllS \right)
 -  \frac{\Ql}{\ClS}  \Biggr]
             \nll
 && +\left( S^2 \yh + 2\Mp \Qh \right)
             \nll
 &&\times   \Biggl[ 2 \Me \Qh \BCll
 -  \frac{\Ql \Qh}{\Qlh}\left(\lClS -\lCllS \right)
 -  \frac{\Qh}{\CllS}
 -  \frac{2}{\SLQ}\Biggr]
             \nll
 && +S^2\yh \left[ \frac{\Qh}{\ClS }
                        +  \frac{\Ql}{\CllS}\right]
         - 4\Mp\frac{B_2}{\lambda^{3/2}_q}
                                   \Biggr\},
\ea
and
\ba
{\cal S}^{L}_{v1}(\yl,\Ql,\yh, \Qh) &=&
-\frac{\yl}{S\yh} \Biggl\{\Biggl(S^2\yl+2\Mp\Ql\Biggr)\Biggl
(\Qlh-\ylh\Qh\Biggr)
\nll &&
            -S^2\ylh\Biggl[\Ql-\left(1-\ylh\right)\Qh\Biggr]
\Biggr\},
\\ \nll
{\cal S}^{L}_{a1}(\yl,\Ql,\yh, \Qh) &=&
-\frac{\yl}{S\yh} \Biggl\{\Biggl(S^2\yl+2\Mp\Ql\Biggr)\Biggl
(\Qlh-\ylh\Qh\Biggr)
\nll &&
            +S^2\Biggl[\ylh\Biggl(\Ql+\left(1-\ylh\right)\Qh
\Biggr)-2\Qlh\Biggr]
\Biggr\},
\\ \nll
{\cal S}^{L}_{v2}(\yl,\Ql,\yh, \Qh) &=&
- 4 \frac{\yl}{S\yhS}\Mp \Ql \ylhS \Qh,
\\ \nll
{\cal S}^{L}_{a2}(\yl,\Ql,\yh, \Qh) &=&
- 4 \frac{\yl}{S\yhS}\Mp
\Ql\Biggl(\Qlh-\ylh\Ql+\ylhS\Qh\Biggr),
\\ \nll
{\cal S}^{L}_{\chi 3}(\yl,\Ql,\yh, \Qh) &=&
\frac{\yl}{S^2\yhS} {\Mp} \Biggl\{
            S^2 \Ql \ylhS \Biggl[ \yll+3 \left(1-\ylh \right
) \Biggr]
\nll
&&           -\Biggl(\Qlh-\ylh\Qh\Biggr)
  \Biggl[S^2\yh\Biggl(2+\left(1-\ylh\right)\frac{\Ql}{\Qh}
\Biggr)
+4\Mp\Ql\Biggr]
                          \Biggr\},
\\ \nll
{\cal S}^{L}_{\chi 4}(\yl,\Ql,\yh, \Qh) &=&
\frac{\yl}{S^2\yhS} \Biggl(S^2 \yh + 2 \Mp \Ql \Biggr) \Biggl
\{
               -S^2\ylhS\left(2\yll+\yh\right)
\nll
&&             +\Biggl(\Qlh-\ylh\Qh\Biggr)
           \Biggl[S^2\yh\left(1-\ylh\right)\frac{1}{\Qh}+2\Mp
\Biggr]
                                                 \Biggr\},
\\ \nll
{\cal S}^{L}_{\chi 5}(\yl,\Ql,\yh, \Qh) &=&
-2\frac{\yl}{S \yh}\Biggl(S^2\yh+2\Mp\Ql\Biggr)
                     \Biggl(\Qlh-\ylh\Qh\Biggr).
\ea
The combined structure functions ${\cal G}^{v,a,\chi}_i(x,Q^2)$
read
\ba
{\cal G}^v_{1,2}(x,Q^2) &=& \lambda_l \left[
Q^2_{e}g^{\gamma \gamma}_{1,2}(x,Q^2)
            + 2|\Qe|v_e\chi(Q^2) g^{\gamma Z}_{1,2}(x,Q^2)
              + v^2_e\chi^2(Q^2)g^{ZZ}_{1,2}(x,Q^2) \right], 
            \nll
{\cal G}^a_{1,2}(x,Q^2) &=& \lambda_l
a^2_e\chi^2(Q^2)g^{ZZ}_{1,2}(x,Q^2),  \nll
{\cal G}^{\chi}_{3,4,5}(x,Q^2) &=& \lambda_l \left [
|\Qe|a_e\chi(Q^2) g^{\gamma Z}_{3,4,5}(x,Q^2)
              + v_e a_e \chi^2(Q^2) g^{ZZ}_{3,4,5}(x,Q^2) \right]~.
\label{addgenstf}
\ea
\subsubsection{Transverse Polarization}
\label{sect323}

\vspace{1mm}
\noindent
Here the differential Bremsstrahlung contribution is given by
\ba
\frac{\ds d\sigma^{{\mr {Trans}}}_{{\mr {Brem}}}}{\ds d\xl
d\yl}
&=&  2  \alpha^3
     \lambda_N^T \COSPHI \;  \frac {d\varphi}{2\pi}
\sqrt{\frac{4\Mp \xl}{S\yl}\left(1-\yl-\frac{\Mp \xl \yl }{S
}\right)}  \nll
&& \times \int d\yh d\Qh
\Biggr\{
\frac{1}{\QhS}
     \sum_{i=1}^{5} \; S^{T}_{i}(\yl,\Ql,\yh,\Qh) {\cal G}_i(\xh,\Qh)
\nll
& &~~~~-\frac{1}{Q^4_l}
\sum_{i=1}^{5}\;S^{T}_{i}(\yl,\Ql)
{\cal L}^{\rm IR}(y_l,Q^2_l,y_h,Q^2_h)
{\cal G}_i(\xh,\Qh) \nll
& &~~~~+ 2  \frac{\Me}{Q_h^4}
 \frac{1}{{C_1}^{3/2}} \Biggl[
 \sum_{i=1}^{2}\Biggl(
   {\cal S}^{T}_{vi}(\yl,\Ql,\yh,\Qh){\cal G}^v_i(\xh,\Qh)
  +{\cal S}^{T}_{ai}(\yl,\Ql,\yh,\Qh){\cal G}^a_i(\xh,\Qh)\Biggr)
\nll
& &~~~~+\sum_{i=3}^{5}
   {\cal S}^{T}_{\chi i}(\yl,\Ql,\yh,\Qh){\cal G}^{\chi}_i(\xh,\Qh)
\Biggr]
\Biggr\}~.
\ea
The corresponding coefficients
$S^{T}_{i}(\yl,\Ql,\yh, \Qh)$,
${\cal S}^{T}_{vi,ai}(\yl,\Ql,\yh, \Qh)$, and
${\cal S}^{T}_{\chi i}(\yl,\Ql,\yh, \Qh)$, are~:
 \ba
 S^{T}_{1}(\yl,\Ql,\yh,\Qh) &=&
      S{\frac{\ylS}{\yh}} \Biggl\{
      \frac{S^2\yl\ylh+2\Mp\Qlh}{\SLQS}                 \\
 && \times \Biggl[-2\Me\Qh\left(\BCl-\BCll\right)
    -(\Ql+\Qh)\left(\lClS+\lCllS\right)   \Biggr]
            \nll
 && - 2\Me\Qh\Biggl(\BCl+\BCll\Biggr)
    +\Qh\frac{\Ql+\Qh}{\Qlh}\left(\lClS-\lCllS\right)
         \nll
 &&      + \Ql\left(\lClS+\lCllS\right)
            \nll
 && +2 \SQSK\frac{S^2\yl+2\Mp\Ql}{\SLQS}
            \nll
 && \times
   \Biggl[ 2\Me\Qh \left(\lCl-\lCll\right)
            \nll
 &&  + (\Ql+\Qh)\left(\FCZl +\FCZll\right) \Biggr]
 \Biggr\},
            \nonumber
\ea
\ba
S^{T}_{2}(\yl,\Ql,\yh,\Qh)&=&
                  { 2S \frac{\ylS}{\yhS} }
   \Biggl\{
   \Qh \left[ 1-\ylh-\yll\frac{S^2\yl\ylh+2\Mp \Qlh}{\SLQS}\right] \\
   &&\times \left[-2\Me  \BCl
     + \frac{\Ql}{\Qlh}\left( \lClS - \lCllS \right)-\lClS\right
]    \nll
   && +\Qh \left[1-\frac{ S^2\yl\ylh+2\Mp \Qlh}{\SLQS} \right
]     \nll
   &&\times
 \left[- 2\Me  \BCll
        +   \frac{\Ql}{\Qlh} \left( \lClS - \lCllS \right)
        +\lCllS  \right]
           \nll
   &&   +   2 \SQSK \Qh
         \frac{S^2\yl+2\Mp \Ql}{\SLQS} \nll
   && \times
 \Biggl[- 2 \Me \left( \frac{\yll}{\ClSS}+\lCll \right)
           \nll   &&   + \FCZl  - \yllFCZll
                          \nll
   &&   + \frac{(2-\yl)\Qh}{\Qlh} \left( \FCZl-\FCZll \right)
 \Biggr]
   \Biggr\},
           \nonumber
\ea
\ba
  S^{T}_{3}(\yl,\Ql,\yh,\Qh) &=&
     \frac{\ylS}{\yhS} \Biggl\{
          \frac{ S^2\yh \ylh+2\Mp \Qlh}{\SLQS}
      \\
     &&\times     \Biggl[
         -\Me\Biggl( \left[S^2 \yll (2\yll + \yh) -2\Mp \Qh\right
] \BCl \nll\
     &&  + \left[S^2      (2     - \yh) -2\Mp \Qh\right] \BCll
\Biggr)  \nll
     &&   +\left[S^2 \Qh \left( \ylpl - \frac{\yl\yh}{2} \right)
           - \Mp (\QlS+\QhS)\right]
           \frac{1}{\Qlh} \left(\lClS - \lCllS  \right)
         \nll
     &&    -\frac{S^2 \yh}{2}\left( \lClS+\frac{\yll}{\CllS}
\right)
          -\frac{2\Mp}{\SLQ} \Biggr]
           \nll
     &&    + \Me\Biggl( \left[S^2 (1-\ylh) (2\yll+\yh)
                     -2\Mp \Qh\right]     \BCl
          \nll
     &&   +\left[S^2       (2     -\yh)-2 \Mp \Qh\right] \BCll
\Biggr)   \nll
     &&   -\Biggl[S^2 \Qh \Biggl(\ylpl
           +\frac{(3\yll-\yhl) \yh}{2}\Biggr)
           -\frac{\Mp (\QlS+\QhS)}{\Qlh}
            \left(\lClS - \lCllS  \right) \Biggr]
          \nll
     &&    +\frac{S^2 \yh}{2} \left(\lClS
           +\frac{1-\ylh}{\CllS}\right)+\frac{2\Mp}{\SLQ}
          \nll
     &&+ \SQSK  \frac{S^2\yl+2\Mp\Ql}{\SLQS }
          \nll     &&    \Biggl[
          2\Me \Biggl( \left[S^2\yll(2\yll+\yh)-2\Mp\Qh\right]
   \lCl  \nll
     &&  +  \left[S^2      (2     -\yh)-2\Mp\Qh\right] \lCll
\Biggr)    \nll
     &&    -\Bigl[S^2 \Qh \left(2 \ylpl-\yl\yh\right)
                        -2 \Mp (\QlS+\QhS)\Bigr]
\nll && \times
                                                 \FCZ
           \nll
     &&    +S^2 \yh \Biggl(\FCZl
                  +\yllFCZll \Biggr) \Biggr]
                                               \Biggr\},
           \nonumber
\ea
\ba
  S^{T}_{4}(\yl,\Ql,\yh,\Qh) &=&
            { \frac{\ylS}{\yhS} }  \Biggl\{
        \Biggl[1-\frac{S^2\yl\ylh+2\Mp \Qlh}{\SLQS} \Biggr]
 \Biggl[
          -2 \Me \Biggl( \left[S^2 \yll (1-\ylh )-\Mp \Qh\right]
                   \BCl
        \nll
 &&               +      \left[S^2  \yhl  - \Mp \Qh\right]
                   \BCll  \Biggr)
        \nll
 &&         + \left[S^2 \Qh(\ylpl-\yl\yh) - \Mp(\QlS+\QhS)\right
]
            \frac{1}{\Qlh}
              \left( \lClS -\lCllS  \right)
     \nll
 &&          -S^2\yh\left(  \lClS  +\frac{\yll}{\CllS} \right)
  -\frac{2 \Mp}{\SLQ}
\Biggr]
     \nll
&& + 2\SQSK\frac{S^2\yl+2\Mp\Ql} {\SLQS}\nll
&& \times \Biggl[
       -2\Me \Biggl(
     \left[S^2 \yll (1-\ylh)   -\Mp \Qh\right]  \lCl
     \\
&&  +\left[S^2  \yhl           -\Mp \Qh\right] \lCll \Biggr)
     \nll
&&  +\left[S^2 \Qh(\ylpl -\yl\yh) -\Mp(\QlS+\QhS) \right]
     \nll
&& \times  \FCZ
     \nll
&&  -S^2\yh\left(\FCZl
    +\yllFCZll\right)  \Biggr]
                                 \Biggr\},
     \nonumber
\ea
\ba
S^{T}_{5}(\yl,\Ql,\yh,\Qh) &=&
                   S{ \frac{\ylS}{\yh} } \Biggl\{
      \Biggl[1-\frac{S^2 \yl \ylh + 2\Mp \Qlh}{\SLQS }\Biggr]
    \Biggl[- 2\Me \Qh   \left( \BCl+\BCll \right) \\
   &&   + \frac{\QlS+\QhS}{\Qlh}\left( \lClS-\lCllS \right)
          +  \frac{2}{\SLQ} \Biggr]
     \nll
   &&   +  2 \SQSK\frac{S^2\yl+2\Mp \Ql}{\SLQS}
             \Biggl[-2\Me \Qh \left(\lCl+\lCll\right)
     \nll
   &&   +  \frac{\QlS+\QhS}{\Qlh}
 \left( \FCZl -\FCZll \right)  \Biggr]
 \Biggr\},
     \nonumber
\ea
and
\ba
{\cal S}^{T}_{v1}(\yl,\Ql,\yh, \Qh) &=&
S\frac{\ylS}{\yh}
\Biggl[ \Biggl( \Qlh-\ylh\Qh \Biggr) \BZl
       -\Biggl(\Ql + \Qh + \ylh \Qh \Biggr){\cal B}_1 \Biggr],
\\ \nll
{\cal S}^{^T}_{a1}(\yl,\Ql,\yh, \Qh) &=&
S\frac{\ylS}{\yh}
\Biggl( \Qlh-\ylh\Qh \Biggr)
\Biggl( \BZl + {\cal B}_1 \Biggr),
\\ \nll
{\cal S}^{T}_{v2}(\yl,\Ql,\yh, \Qh) &=&
 2S\frac{\ylS}{\yhS} \Qh
\Biggl[ \ylhS\BZl - \left(\yl - \yll \ylh\right) {\cal B}_1 \Biggr],
\\ \nll
{\cal S}^{T}_{a2}(\yl,\Ql,\yh, \Qh) &=&
 2S\frac{\ylS}{\yhS}
\Biggl(\Qlh-\ylh\Qh\Biggr)
\Biggl[\left(1 - \ylh\right)\BZl - \yll {\cal B}_1 \Biggr],
\\ \nll
{\cal S}^{T}_{\chi 3}(\yl,\Ql,\yh, \Qh) &=&
 \frac{\ylS}{\yhS}\frac{1}{2}
\Biggl\{
\Biggl[-S^2\ylhS\left[\yll+3\left(1-\ylh\right)\right]
\nll
&&\hspace{1cm}
  +\Biggl(\Qlh-\ylh\Qh\Biggr)
   \Biggl(S^2\yh\frac{1-\ylh}{\Qh}+4\Mp\Biggr)
\Biggr] \BZl
\nll
&&\hspace{7mm}
  +\Biggl[S^2
   \Biggl(\yl\left(2-\yl\right)-\left(1+\yllS\right)\ylh+3\yll
\ylhS\Biggr)
\nll
&&\hspace{1cm}
  -\Biggl(\Qlh-\ylh\Qh\Biggr)
   \Biggl(S^2\yh\frac{\yll}{\Qh}+4\Mp\Biggr)
\Biggr] {\cal B}_1
\Biggr\},
\nll \\
{\cal S}^{T}_{\chi 4}(\yl,\Ql,\yh, \Qh) &=&
\frac{\ylS}{\yhS}
\Biggl[S^2\ylhS\left(2\yll+\yh\right)
\nll
&&\hspace{7mm}
  -\Biggl(\Qlh-\ylh\Qh\Biggr)
   \Biggl(S^2\yh\frac{1-\ylh}{\Qh}+2\Mp\Biggr)
\Biggr]
  \Biggl(\BZl - {\cal B}_1
\Biggr),
\\ \nll
{\cal S}^{T}_{\chi 5}(\yl,\Ql,\yh, \Qh) &=&
 2S\frac{\ylS}{\yh}
\Biggl(  \Qlh - \ylh \Qh  \Biggr)
\Biggl( \BZl - {\cal B}_1
\Biggr)~.
\ea
The quantity ${\cal B}_1$, and similarly ${\cal B}_2$, which
contain  the
denominator $\lambda_q$, may be simplified in the following way for
the $O(m^2)$ terms,
\ba
 {\cal B}_1 &=& \BZl \frac{S^2 \yl\ylh + 2 \Mp \Qlh}{\SLQS}
     - 2 \SQSK \frac{S^2 \yl     + 2 \Mp \Ql} {\SLQS}
 \nonumber\\
&\equiv& S^2 y_{lh} (Q^2_{lh}-y_{lh} Q^2_l),\\
 {\cal B}_2 &=& B_2  \frac{S^2 \yl\ylh + 2 \Mp \Qlh}{\SLQS}
     - 2 \SQSK \frac{S^2 \yl     + 2 \Mp \Ql} {\SLQS}
 \nonumber\\
&\equiv& S^2 y_{lh} Q^2_{lh} (1 + y_h) - S^2 y_{lh}^2 Q^2_{h}
+ 2 M^2 (\Qlh)^2.
\ea
The explicit substitution of these terms in the above relations does
not lead to a further simplification.
%
\section{The Leading Log Approximation}
\label{sect4}

\vspace{1mm}
\noindent
In the leading logarithmic approximation the leptonic
$O(\alpha)$  corrections  to the differential scattering cross section
can be written as
\ba
\frac{d^2 \sigma_{\mr {rad}}^{\rm LLA}}{dxdy}
&=&
\frac{d^2 \sigma_{i}}{dxdy}
+
\frac{d^2 \sigma_{f}}{dxdy}
+
\frac{d^2 \sigma_{\rm{Comp}}}{dxdy}.
\label{lla3}
\ea
The individual contributions are due to the corresponding
mass-singularities of initial-state
({\it i}) and final-state ({\it f}) radiation, and the Compton
contribution, for which the mass singularity results from the
low $Q^2$
range of the virtual photon in the sub-system Born term, in those
situations which may be described partonically.

The initial and final-state radiation terms have the following
structure:
\ba
\frac{d^2 \sigma^{k}_{i,f}}{dxdy}
=
\frac{\alpha}{2\pi}
\left( \ln\frac{Q^2}{m^2}-1 \right)
\int\limits_{0}^{1}dz
\frac{1+z^2}{1-z}
\Biggl\{\theta(z-z_0)~{\cal J}\left.
\frac{d^2
  \sigma^{k}_{{\rm{Born}}}}{dxdy}\right|^{i,f}_{x={\hat x},
 y={\hat y}, S={\hat S}}
-\frac{d^2 \sigma^{k}_{{\rm{Born}}}}{dxdy}\Biggr\},
\label{LLA1}
\ea
where the rescaled variables are
\ba
\hat{S} = zS,~~~\hat{y} = \frac{y+z-1}{z},~~~\hat{Q^2} = zQ^2,~~~{\hat x}
 &=& \frac{{\hat Q}^2}{{\hat y}{\hat S}},
\ea
for initial-state radiation and
\ba
\hat{S} = S,~~~\hat{y} = \frac{y+z-1}{z},~~~\hat{Q^2} = 
\frac{Q^2}{z},~~~{\hat x}
 &=& \frac{{\hat Q}^2}{{\hat y}{\hat S}},
\ea
for final-state radiation.
The Jacobian ${\cal J}$
is given by
\ba
{\cal J} \; \equiv \; {\cal J}(x,y,Q^2) =
\left|\frac{\partial({\hat x},{\hat y})}{\partial(x,y)}\right|,
\label{llaif}
\ea
cf.~refs.~\cite{HADR1,lla1}--\cite{lla2,llA1}.
The lower integration boundaries  $z_0$ derive  from the conditions
\ba
{\hat x}(z_0) &\leq& 1, \qquad
{\hat y}(z_0) \; \leq\;  1,
\label{llascal}
\ea
and are given by
\ba
z_0^i =  \frac{1-y}{1-y x},~~~~~~~~~~~~~~~~~~~~z_0^f  = 1 - y + x y~.
\ea
The structure of eq.~(\ref{LLA1}) is the same in the unpolarized and
polarized case in leading order QED, since, as is
well--known~\cite{SPLI}, the fermion--fermion splitting function
\ba
P_{ff}(z) = \left ( \frac{1 + z^2}{1 - z} \right )_+
\ea
does not depend on the polarization of the incoming fermion in this
order. Therefore, the Bremsstrahlung contributions for the polarized
case are simply obtained in LLA by  inserting in
the
relations~refs.~\cite{HADR1,lla1}--\cite{lla2,llA1}
the differential Born cross section for the polarized case
\cite{lla3}.
Similar relations were also calculated in $O(\alpha^2 \ln^2(Q^2/m^2))$
in refs.~\cite{kripf,JB2}.

For leptonic variables the Compton
contributions depend on the type of nucleon polarization. These
contributions are given by
\ba
\frac{d^2 \sigma_{{\rm{Comp}}}^{U}}{d\xl d\yl}
&=&
\frac{\alpha^3}{Sx^2_l } \frac{Y_{+}}{y_{l_1}}
  \int \limits_{x_l}^{1} d z
  \int\limits^{(Q^2_h)^{\rm{\small{max}}}}_{(Q^2_h)^{\rm{\small{min}}}}
  \hspace{-2mm}
  \frac{d\Qh}{\Qh}
  \Biggl[\frac{Z_{+}}{z} F^{\gamma\gamma}_2(\xh,\Qh)
-      z F^{\gamma\gamma}_{L}(\xh,\Qh)\Biggr],~~~~{\rm
cf.~\cite{lla4,lla2,lla1,coU}},
\label{compt_u}
\nll
\frac{d^2 \sigma_{\rm Comp}^{L}}{d\xl d\yl}
&=&
  \left(- 2 \lambda_l \lambda^{L}_N \right)
  \frac{\alpha^3}{S x^2_l }\frac{Y_-}{\yll}
  \int \limits_{x_l}^{1} {d z }
  \int\limits^{(Q^2_h)^{\rm{\small{max}}}}_{(Q^2_h)^{\rm{\small{min}}}}
\hspace{-2mm}
  \frac{d\Qh}{\Qh} \frac{Z_{-}}{z} x_h g^{\gamma\gamma}_1(\xh,\Qh),
\nll
\frac{d^2 \sigma_{{\rm{Comp}}}^{T}}{d\xl d\yl}
&=&
  \left(- 2 \lambda_l \lambda^{T}_N \right)
  \frac{\alpha^3}{S x^2_l }
  \cos \varphi \frac{d\varphi}{2 \pi}
  \frac{y_l}{y^2_{l_1}}
  \sqrt{\frac{4 M^2 x_l}{S y_l} \left(\yll-\frac{M^2 x_l y_l}{S} \right)}
\nll
&\times&
\int \limits_{x_l}^{1} \frac{dz}{z}
\int\limits^{(Q^2_h)^{\rm{\small{max}}}}_{(Q^2_h)^{\rm{\small{min}}}}
\hspace{-2mm}
  \frac{d\Qh}{\Qh}
  \Biggl\{
  \left( Y_{-} - y_l z \right) z
 x_h g^{\gamma\gamma}_1(\xh,\Qh)
+ 2  \left[ Y_{+}\left(1-z \right)
 + y_{l_1} \right]
  x_h g^{\gamma\gamma}_2(\xh,\Qh)
  \Biggr\}~.
\nonumber\\
\label{compt}
\ea
Here we used the abbreviations
\ba
\ylpl &=& 1+(1-\yl)^2,~~~~~~~~~~~~~~~~~~~~~Z_{\pm} = 1 \pm (1-z)^2,
\ea
\ba
z\;=\;\frac{\xl}{\xh}.
\label{yz}
\ea
Part of the structure of eqs.~(\ref{compt}) can be understood, as in the
case of Bremsstrahlung, in
terms of partonic splitting functions. In the unpolarized case
\ba
P_{\gamma f}^{U}(z) = \frac{Z_+}{z} = \frac{1 + (1-z)^2}{z}
\ea
emerges~\cite{lla2,llA1}.
The longitudinal structure function is convoluted by the coefficient
function
\ba
c_L^q(z) = z,
\ea
cf.~\cite{TWZ}.
In the case of longitudinal nucleon polarization one has~\cite{SPLI}
\ba
P_{\gamma f}^{L}(z) = \frac{Z_-}{z} = \frac{1 - (1-z)^2}{z},
\ea
and the Compton term can as well be understood in the collinear parton
model. Such an interpretation is, however, not possible in the case of
transverse nucleon polarization, see~ref.~\cite{BK}.

For the boundaries of the $\Qh$-integral in eqs.~(\ref{compt}) we  used
\ba
(Q^2_h)^{\rm{\small{max}}} &=& \frac{\xl}{\xh}(1-\yl)\Ql,
\nll \nll
(Q^2_h)^{\rm{\small{min}}} &=&
\max
\left\{
(Q^2_h)^{\rm{\small{min}}}_{\rm{kin}},\overline{Q}^2_h,{\hat Q}^2_h
\right\},
\label{q2lim}
\ea
where
\ba
{\hat Q}^2_h &=& \left(\overline{M}^2_h-\Mp\right)\frac{\xh}{1-\xh},
\nll
(\Qh)_{\rm kin}^{\rm min} &=& \frac{\xh(\xh \yl S - \Ql)(\yl S
-  \sqrt{\lambda_q}) + 2 \Ql (\xh M)^2}{2\left [ \xh \yl S - \Ql
+ (\xh M)^2 \right ]}.
\ea
$\overline{Q}^2_h$ and $\overline{M}^2_h$ denote respective cut values.
The choosen value of $(Q^2_h)^{\rm max}$ leads to a particularly good
agreement of the LLA results and the complete $O(\alpha)$ corrections.

The LLA formulae are remarkably compact. The initial and final--state
radiation contributions rely only on the respective Born cross sections
in rescaled kinematic variables.
Also the Compton terms exhibit a similarly simple structure.
A numerical comparison of the complete leptonic
$O(\alpha)$ corrections and the corresponding LLA results is  given in
section~5.
%
\section{Numerical Results}
\label{sect5}

\vspace{1mm}
\noindent
The QED corrections, which are discussed subsequently, are presented
in terms of the correction factors
\ba
\delta = \frac{d^2 \sigma^{\rm pol}_{\rm rad} }
              {d^2 \sigma^{\rm pol}_{\rm Born}} \times 100 \%
\ea
for the different nucleon polarizations and for different kinematic
ranges. The calculations were performed with the code 
{\tt HECTOR}~\cite{HEC2}. For the parametrization of the polarized
parton densities  we use the `standard' set (LO) of ref.~\cite{GRSV}.
The non-perturbative low $Q^2$-behavior of the polarized parton densities
is modeled following an  approach~\cite{PROKH} which was used
in ref.~\cite{R3} in  the unpolarized case
before.

In Figure~5 the complete leptonic $O(\alpha)$  QED correction to the
polarized part of the differential deep-inelastic scattering cross
section for longitudinally
polarized protons is shown for the kinematic   range of the HERMES
experiment at HERA at an electron beam energy of $E_e = 27.5 \GeV$.
Similarly to the case of unpolarized deep-inelastic
scattering, see e.g.~\cite{R1,R3},
the corrections rise towards small $x$ and high
$y$ values. The correction was calculated for the case of pure
photon exchange. The structure functions were parametrized referring to
the twist 2 contributions,
using the parton densities of ref.~\cite{GRSV}.
The complete $O(\alpha)$ corrections are compared with the corresponding
LLA results. A very good agreement of both descriptions is obtained
for larger values of $x$ and lower and medium values of $y$. In the
high~$y$ range the LLA contributions yield  somewhat larger corrections,
which are further enhanced by the Compton contribution.

In Figure~6 the same corrections are considered applying a cut of
$Q^2_h > 1 \GeV^2$. The Compton part of the LLA contributions is
widely removed by this condition. For smaller $x$ values the
agreement between the complete corrections and the LLA contributions
becomes somewhat worse.

Figure~7 shows the different contributions to the leptonic corrections
in LLA for the kinematic regime of the HERMES experiment for longitudinal
proton polarization. While the
initial state corrections behave rising for growing $y$ and smaller
values of $x$, the final state corrections are rather flat in $y$.
The Compton contribution is only essential at small $x$ and large $y$.

In Figure~8 a comparison is given for the complete leptonic
$O(\alpha)$ corrections and the corresponding LLA contributions for
the polarized part of the
differential scattering cross section in the case of transverse
proton polarization for the kinematic range of the HERMES experiment.
The $x$-dependence of the corrections is implied by the behavior
of the structure functions $g_1(x,Q^2)$ and $g_2(x,Q^2)$.
As in the case of longitudinally polarized protons,
 the complete and
LLA corrections agree very well for small $y$ and larger values of $x$.
In the high $y$ range both descriptions differ. Again the Compton
term increases the difference slightly.

The effect of a cut of $Q^2_h > 1 \GeV^2$ on the corrections for
the case of transverse proton polarization is illustrated in Figure~9.
The Compton contribution is removed by this cut as in the case
of longitudinal proton polarization (Figure~6). For smaller values
of $x$ the LLA terms deviate stronger from the complete corrections
than in the case of longitudinal proton polarization.

In Figure~10 the different contributions to the LLA corrections
are depicted for the case of transverse nucleon polarization.
Similarly to the case of longitudinal nucleon polarization, the
initial state radiation terms do widely determine the structure of the
LLA correction. The final state radiation terms are flat in $y$ and
the Compton terms contribute at small $x$ and large $y$ only.

The polarized contribution to the
complete leptonic $O(\alpha)$ correction to the
differential deep-inelastic scattering cross section for longitudinally
polarized protons in the kinematic range of the HERA collider
at $\sqrt{S} = 314~\GeV$ are shown in Figure~11.
Their shape is rather similar to the corrections for lower values of
$\sqrt{S}$, cf.~Figure~5. Lower values of $x$, $x \sim 10^{-3}$ can
be probed. Here the corrections are rather large. The comparison with
the LLA contributions is also shown. While the full LLA corrections
agree rather well with the results of the complete calculation in the
whole kinematic range, the contributions due to only initial and final
state Bremsstrahlung differ in the high $y$ and low $x$ range from
the complete $O(\alpha)$ corrections. The good
agreement of the complete LLA relations is due to the choice of the
Compton-logarithms, eq.~(\ref{compt},\ref{q2lim}), which is particularly
suited for larger values of $\sqrt{S}$.

In Figure~12 a comparison is shown of the polarized part of the
complete leptonic $O(\alpha)$  QED corrections for longitudinally
polarized protons at $\sqrt{S} = 314~\GeV$ parametrizing the respective
contributions to the polarized part of the scattering cross section
by either only the structure function $g_1(x,Q^2)$ or the complete
set of structure functions $g_1(x,Q^2), g_4(x,Q^2)$, and $g_5(x,Q^2)$,
which are not  suppressed kinematically
by factors of $O(M^2/S)$, as the contributions due to the
structure functions
$g_2(x,Q^2)$ and $g_3(x,Q^2)$. Because of
the different $y$ behavior of
the kinematic factors weighting the structure functions,
the correction
changes significantly if only the structure function $g_1(x,Q^2)$ is
considered in the scattering cross section. This applies also for the
small $x$ and low $y$ range and is caused by the structure functions
$g_4(x,Q^2)$ and $g_5(x,Q^2)$,  which contribute to the $\gamma$-$Z$
interference and $Z$-exchange contributions.
%
\section{Conclusions}
\label{sect6}

\vspace{1mm}
\noindent
The $O(\alpha)$ leptonic QED corrections to the polarized
differential neutral
current  deep-inelastic scattering cross sections were calculated
both for the case of longitudinal and transverse nucleon polarization.
This approach allows for a general description of
structure functions, and is therefore particularly suited for
experimental analyses, which aim on the unfolding of the different
polarized
structure functions from the measured cross sections.
In the explicit calculations we considered a hadronic tensor which obeys
Lorentz and time--reversal invariance, as well as current conservation.
Its polarized part is described by the five structure functions 
$\left. g_i^p(x,Q^2)\right|_{i=1}^5$.
In the $O(\alpha)$ corrections the general structure of the Born
cross section is resembled. The corrections can be expressed in terms
of general kinematic factors and appropriate combinations of the
neutral current structure functions, including both photon and
$Z$-boson exchange.
Additional combinations of
structure functions contribute as well for terms $\propto m^2$.
The latter terms yield finite contributions after integration in
the limit $m \rightarrow 0$.

In the leading logarithmic approximation the $O(\alpha)$ leptonic
QED corrections are described by the initial and final state
Bremsstrahlung contributions and the Compton terms for pure
photon exchange. While the structure of the Bremsstrahlung contributions
is the same for the case of unpolarized and polarized deep-inelastic
scattering, different expressions are obtained for the Compton terms.
This is due to the behavior of the leading order QED
splitting and coefficient functions in situations, in which the
collinear parton model applies. The respective expressions are rather
compact and their structure is partly induced by  mass
singularities, which allows an easy deduction of the corresponding
expressions.

Numerically the $O(\alpha)$ corrections are rather large in the
case of $e N$-scattering. Since the corrections behave 
$\sim \ln(Q^2/m^2)$,
smaller corrections are obtained for the case of polarized $\mu N$
scattering. 
A numerical comparison with the corrections obtained by the
code~\cite{MIN1} was carried out in refs.~\cite{HWS}--\cite{CRAC}
for the simplified assumptions in~\cite{MIN1}.

Both in the kinematic domain
probed by the HERMES experiment and the
kinematic range which would be
accessible in
possible future polarized
deep inelastic scattering experiments in the HERA collider mode,
the corrections grow towards small $x$ and high $y$.
Except the range of large values of $y$ and small $x$ already
the LLA results provide a very good description of the corrections. 
The structure of the corrections as a function of $x$ and $y$ is  widely
determined by the initial-state Bremsstrahlung terms.
If cuts, e.g.
on $Q^2_h$, are applied this agreement, however,
becomes worse.
While for longitudinal nucleon polarization the corrections are 
structurally similar to those in the
unpolarized case, the corrections
in the case of transverse nucleon polarization show a different
$x$-behavior.
%
\newpage
\setcounter{equation}{0}
\appendix
\def\theequation
{\Alph{section}
.\arabic{equation}}
\section
{Kinematic   relations for the Bremsstrahlung process
and an analytical integral
\label{sectA}           }

\vspace{1mm}
\noindent
We summarize a series of
kinematic relations for the Bremsstrahlung process
which are used in the present calculation. In the rest frame of the
nucleon ($\vec p =0 $) the proton polarization 4--vector $s$ and
the lepton 4--vectors are given by
\ba
 s &=& M  (0,\cos\varphi,\sin\varphi,0),    \nll
 k_1 &=& (k_{10},0,0,|\vec k_1|),             \nll
 k_2 &=& (k_{20},|\vec k_2| \sin\theta_2,0,|\vec k_2|\cos\theta_2),
\ea
with
\ba
 q_l &=&k_1-k_2
      =(k_{10}-k_{20},-|\vec k_2|\sin\theta_2,0,|\vec k_1|
-|\vec k_2|\cos\theta_2).
\ea
\begin{figure}[bhtp]
\begin{minipage}[bhtp]{14.5cm}{
\begin{center}
\begin{Feynman}{145,60}{-30, 15}{1.5}
\thicklines
\put(-7,30){${\varphi}$}                    
\put(0 ,24  ){\oval( 8,9 )[tl]}             %
\put(00,22.5){\line(-3, 1){15}}             %
\put(-16,28){\vector(-3, 1){0}}             %
\put(-19,26 ){$ s            $}             %
\put(0 ,27.2){\oval(9 ,10)[tr]}             %
\put(3 ,33){${\gamma }$}                    
\put(45,7.5){\line(-3, 1){45}}              
\put(22.5,15){\vector( 3,-1){0}}            %
\put(22,10){${\vec \Lambda}$}               %
\put(45,7.5){\line(1,1){15}}                
\put(52.5,15){\vector( 1, 1){0}}            %
\put(52,09){${\vec p}_2$}                   %
\multiput(00,22.5)(6,0){10}{\line(1,0){4}}  
\put(28,22.5){\vector(1, 0){0}}             %
\put(26,25){${\vec k_1}$}                   %
\put(0,22.5){\line(4, 3){34}}               
\put(26,37){${\vec k}_2$}                   %
\put(25,41.3){\vector(4, 3){0}}             %
\put(32,60.5){\oval(10,10 )[t] }            
\put(27,59  ){\vector(0,-1){0}}             %
\put(24,58  ){$\varphi_k$     }             %
\put(34.5,48){\line(1,-1){25}}              
\put(45,37.5){\vector( 1,-1){0}}            %
\put(46,40){${\vec Q}_l$}                   %
    \put(39,37){${\theta_k}$}
    \put(37.0,42){\oval(6.5,6)[br]}
\put(45,7.5){\line(-1,4){10}}               
\put(40,28){\vector( 1,-4){0}}              %
\put(35,28){${\vec k}  $}                   %
%
\put(34.5,48){\circle*{1.5}}
\put(00,22.5){\circle*{1.5}}
\put(45,7.5){\circle*{1.5}}
\put(60,22.5){\circle*{1.5}}
\thinlines
\put(00,22.5){\line(1,1){40} }        
\put(40,62.5){\vector(1,1){0}}        
\put(41,63)  {$  x_{_R}$     }        
\put(00,22.5){\line(0,1){50} }        
\put(-.5,73.5  ){$ x $       }        
\put(-16,28){\line(0,-1){4.5}}
\put(-16,22.5){\line(0,-1){4.5}}
\put(-16,17){\line(0,-1){4.5}}
\put(-16,11.5){\line(0,-1){4.5}}
\put(-16,28){\line(1,1){3}}
\put(-12,32){\line(1,1){3}}
\put(-8 ,36){\line(1,1){3}}
\put(-4 ,40){\line(1,1){3}}
\put( 00,22.5){\line  (-1,-1){25.3}}  
\put(-25.4,-3){\vector(-1,-1){0 }}    %
\put(-27  ,-5){$y,\; y_{_R}        $} %
\put(73,22  ){$z$     }               
\put(60,22.5){\line  (1,0){10}}       %
\put(70,22.5){\vector(1,0){0} }       %
\put(00,72,5){\vector(0,1){0}}        %
\put(0 ,22.5){\line(1,-1){25}}        %
\put(25,-2.5){\vector(1,-1){0}}       %
\put(26,-5){$     z_{_R} $  }       %
\put(45  ,7.5){\line(1,-4){2}}        
\put(34.5,48 ){\line(-1,4){3}}        
\thicklines
\put(7.9  ,22.5){\oval(11 ,11 )[tr]}          
\put(14,26.5){${\theta_2 }$}                    %
\end{Feynman}
\end{center}
}\end{minipage}
\vspace*{3.cm}
\end{figure}

\vspace{2mm}
\noindent
\begin{center}
{\sf Figure~4:
3--vector relations for the Bremsstrahlung process in
the rest frame of the nucleon.}
\end{center}
We will also use  a rotated  frame in which two
momentum components of $q$ vanish,
\ba
 q_{lR }&=&
      = (q_{l0}, 0, 0, |\vec q_l|)
    \nll
 k_R  &=& k_{0R} (1,\sin\theta_k \cos\varphi_k,
                   \sin\theta_k \sin\varphi_k,
                   \cos\theta_k).
\ea
Here
\ba
 k_0     = \frac{S \ylh}{ 2 M  },~~~~k_{10}   =
\frac{S}{2 M  },~~~~k_{20}   = \frac{S \yll}{2 M  },
\ea
and
\ba
\left|\vec k   \right|   = \frac{\SLK}{2 M  },~~~~\left|\vec
k_1 \right|   = \frac{\SLS}{2 M  },~~~~\left|\vec k_2
\right|   = \frac{\SLL}{2 M  }.
\ea
The triangle functions $\lambda_X$ are
\ba
     \LK       &=& S^2 y_{lh}^2         \nll
     \LS       &=& S^2-4 m^2 M^2,            \nll
     \LL       &=& S^2 \yllS - 4 m^2 M^2.
\label{eqLAM1}
\ea
The sine and the
cosine of the rotation angle $\gamma$ are
\ba
 \cos \gamma &=& \frac{|\vec k_1|
- |\vec k_2|\cos\theta_2 }{|\vec Q_l|},\\
 \sin \gamma &=& \frac{|\vec k_2|\sin\theta_2 }{|\vec Q_l|}.
\ea
The transformation between the rest frame and the rotated frame
for a general 4--vector $K$ is given by
\ba
 K_0 &=& K_{0R}, \nll
 K_x &=& K_{xR}\cos\gamma - K_{zR}\sin\gamma, \nll
 K_y &=& K_{yR},                              \nll
 K_z &=& K_{xR}\sin\gamma + K_{zR}\cos\gamma.
\ea
The 4--vector $q$ is thus represented by
\ba
 q = (k_{10}-k_{20}, - |\vec k_2| \sin\theta_2, 0,
|\vec k_1| - |\vec k_2| \cos\theta_2)
\ea
in the nucleon
rest frame. The 4--vector of the radiated photon is given by
\ba
 k_0 &=& k_{0R}, \nll
 k_x &=& k_{0R}\left( \sin\theta_k \cos\varphi_k \cos\gamma
                  - \cos\theta_k  \sin\gamma\right),            \nll
 k_y &=& k_{0R}\sin\theta_k \sin\varphi_k,                     \nll
 k_z &=& k_{0R}\left( \sin\theta_k \cos\varphi_k \sin\gamma
                    + \cos\theta_k \cos\gamma \right).
\label{**}
\ea
The different angles defined in Figure~4 can be expressed in terms of
the invariants (\ref{eqLAM1}, \ref{eqLAM2}) as follows
\ba
 \cos\theta_2 &=&
 \frac{\LS+\LL-\LQ}{2\SLS\SLL}
 \approx  1-\frac{2 M^2 \Ql}{S^2 \yll},          \nll
 \sin\theta_2 &=&
 \frac{\sqrt{-\lambda(\LS,\LL,\LQ)}}{2\SLS\SLL}
 \approx \frac{2 M^2 \SQSll} {S^2 \yll},         \nll
 \cos\gamma   &=&
 \frac{\LS+\LQ-\LL}{2\SLS\SLQ}
\approx \frac{S^2 \yl+2\Mp\Ql}{S \SLQ},          \nll
 \sin\gamma   &=&
 \frac{\sqrt{-\lambda(\LS,\LL,\LQ)}}{2\SLS\SLQ}
 \approx \frac{2 \Mp \SQSll}{S \SLQ},            \nll
 \cos\theta_k &=&
 \frac{\LK+\LQ-\LH}{2\SLK\SLQ}
 \approx \frac{S^2 \yl \ylh + 2\Mp \Qlh}{S \ylh \SLQ}, \nll
 \sin\theta_k &=&
 \frac{\sqrt{-\lambda(\LK,\LL,\LH)}}{2\SLK\SLQ}
\approx \frac{2 \Mp \SQSKS}{S \ylh \SLQ}.
\label{eqANG}
\ea
Here
\be
\lambda(x,y,z) = (x - y - z)^2 - 4 y z
\ee
and
\ba
\label{eqLQ}
     \LQ       &=& S^2 \yl^2+4 M^2 \Ql,      \nll
\label{eqLH}
     \LH       &=& S^2 \yh^2+4 M^2 \Qh,      \nll
\lambda_{Slq}&=&\frac{4\LS\LL-(\LS+\LL-\LQ)^2}{16 M^2}   \nonumber\\
 &\equiv& S^2 \yll
Q_l^2 - M^2 Q^4_l - m^2 \lambda_q,
\nll
  \lambda_{kqh} &=& \frac{4\LK\LQ-(\LQ+\LK-\LH)^2}{16 M^2} \nonumber\\
&\equiv& S^2 y_{lh} d_{lh}- M^2 (Q^2_{lh})^2,
\label{eqLAM2}
\ea
with
\ba
d_{lh} =  y_l Q^2_h - y_h Q^2_l~.
\label{eqdlh}
\ea
The latter terms in eqs.~(\ref{eqANG}) result from  the ultrarelativitic
approximations in the lepton momenta. Furthermore, one obtains
\ba
  \cos\gamma\cos\theta_2
 -\sin\gamma\sin\theta_2 &=&
 \frac{\LS -\LL -\LQ}{2 \SLQ \SLL},                 \nll
  \cos\gamma\sin\theta_2
 +\sin\theta_\gamma\sin\theta_2 &=&
 \frac{\sqrt{-\lambda(\LS,\LL,\LQ)}}{2\SLQ\SLL}.
\ea

The integral over the angle $\varphi_k$ can be performed analytically.
We express the phase space element  as
\ba
 d\Gamma = \frac{\pi^2 S^3}{4\SLS } y_l
           dx_l d\yl d\yh d\Qh \frac{1}{\pi} \frac{dz_i}{\sqrt R_i}
         = \frac{\pi^2 S^3}{4\SLS } y_l
           dx_l d\yl d\yh d\Qh
 \frac{1}{\sqrt{\lambda_q}}\frac{1}{2\pi} d\varphi_k,
\ea
with $i=1$ or $2$, and
\ba
R_i &=& -\lambda_q z^2_i + 2 B_i z_i - C_i
\\
\label{eqB1}
 B_1 &=& S^2\left(\Ql \yll \ylh + d_{lh}\right)-2\Mp
\Ql \Qlh,
\\
\label{eqB2}
 B_2 &=& S^2\left(\Ql \ylh + \yll d_{lh}\right)+2 \Mp \Ql \Qlh,
\\
\label{eqC1}
     C_1 &=& S^2 \left[(\yll+\yh)\Ql-\Qh\right]^2
   + 4 m^2 \left[S^2 \ylh d_{lh} - \Mp (\Qlh)^2 \right],
\\
\label{eqC2}
    C_2&=& S^2 \left[(1-\yh)\Ql-\yll \Qh\right]^2
   + 4 m^2 \left[S^2 \ylh d_{lh} - \Mp (\Qlh)^2 \right]~.
\ea
These quantities are related by the identities
\ba
 B_1 +B_2  &=&  \frac{C_2 -C_1}{\Qlh}, \\
 B_2 -B_1  &=&  \lambda_q \Qlh.
\ea
With the help of the above relations
the invariants $z_1$ and $z_2$, eq.~(\ref{z12}),
can be expressed as
\ba
z_1 &=& a_1 - b \cos \varphi_k, \nll
z_2 &=& a_2 - b \cos \varphi_k,
\ea
where
\ba
a_1  &=& \frac{S\sqrt{\lambda_k}}{2 \Mp}
     -   \frac{ (\Ncg)(\Nck) } {8\Mp \lambda_q},  \nll
a_2  &=& \frac{S\sqrt{\lambda_k}}{2 \Mp} y_{l_1}
     -   \frac{ (\Ncg)(\Nck) } {8\Mp \lambda_q},  \nll
b    &=& \frac{  \NSg  \NSk  } {8\Mp \lambda_q}.
\ea
The different $\varphi_k$-integrals
\ba
\vph
     {\ds \Biggl[ A  \Biggr]_\varphi }
 &= &{\ds \frac{1}{2 \pi}        }
     {\ds \int_0^{2\pi}
        \frac{d\varphi_k }{\sqrt{\lambda_q} }  }~A(\varphi_k)
\ea
are:
\begin{eqnarray}
\begin{array}{lclclcl}
 \vph
      && {\ds \Biggl[  \ZlS \Biggr]_\varphi } &=&
  {\ds \frac{ 3 \BZ- {\lambda_q  }  \CZ }
  { 2{\lambda_q  }^{5/2}}},
\nonumber\\
 \vph
      && {\ds \Biggl[ \frac{1}{z^2_{1(2)}}  \Biggr]_\varphi } &=&
         {\ds \frac{B_{1(2)}}{C_{1(2)}^{3/2}}            },
\nonumber\\
 \vph
      && {\ds \Biggl[ \frac{1}{z _{1(2)}}  \Biggr]_\varphi } &=&
         {\ds \frac{1  }{\SC},          }
\nonumber\\
 \vph
      && {\ds \Biggl[ \frac{1}{ z_1 z_2 }  \Biggr]_\varphi } &=&
    \frac{\ds 1           }{\ds \Qlh}
      {\ds   \Biggl[
 \frac{1}{\sqrt{C_1}}
-\frac{1}{\sqrt{C_2}}
             \Biggr] },
\nonumber\\
 \vph
      && {\ds \Biggl[ z_{1(2)}            \Biggr]_\varphi } &=&
         {\ds \frac{B_{1(2)}}{{\lambda_q  }^{3/2}}       },
\nonumber\\
 \vph
      && {\ds \Biggl[  1 \Biggr]_\varphi }
&=& \frac{\ds 1}{\ds \sqrt{\lambda_q}  },
\end{array}
\end{eqnarray}
\begin{eqnarray}
\begin{array}{lclclcl}
       && {\ds \Biggl[
\frac{\cos \varphi_k}
{ z_{1(2)} }
  \Biggr]_\varphi }
&=& \frac{\ds b  \lambda_q }
         {\ds \sqrt{C_{1(2)}}\Bigl
(B_{1(2)} + \sqrt{\lambda_q} \sqrt{C_{1(2)}
} \Bigr
) },
\nonumber\\
    && {\ds \Biggl[  \frac{\cos \varphi_k}
{ z_{1(2)}^2 }
  \Biggr]_\varphi     }
&=& \frac{\ds b \lambda_q    }
         {\ds C^{3/2}_{1(2)} },
\nonumber\\
    && {\ds \Biggl[  \frac{\cos \varphi_k}{z_1 z_2}   \Biggr]_\varphi}
&=& \frac{\ds b \lambda_q }{\ds \Qlh}
      {\ds   \Biggl[
 \frac{1}{\sqrt{C_1}(B_1+\sqrt\lambda_q \sqrt{C_1}) }
-\frac{1}{\sqrt{C_2}(B_2+\sqrt\lambda_q \sqrt{C_2}) }
         \Biggr] }.
 \end{array}
\end{eqnarray}

\vspace{5mm}
\noindent
{\bf Acknowledgement}~We would like to thank Prof. P. S\"oding for
his constant support of the present project. P.C. and L.K. would like
to thank DESY--Zeuthen for the hospitality extended to them.
D.B. would like to thank I. Akushevich for useful
discussions.
We would like to thank P.~Mulders for providing us ref.~[3b].
The project was partly supported by EC-grant HCMP-CT920004.
and PECO-grant ERBCIPDCT--94--0016.


\vspace{5mm}
\noindent

\normalsize
%
\newpage
\begin{center}

\mbox{\epsfig{file=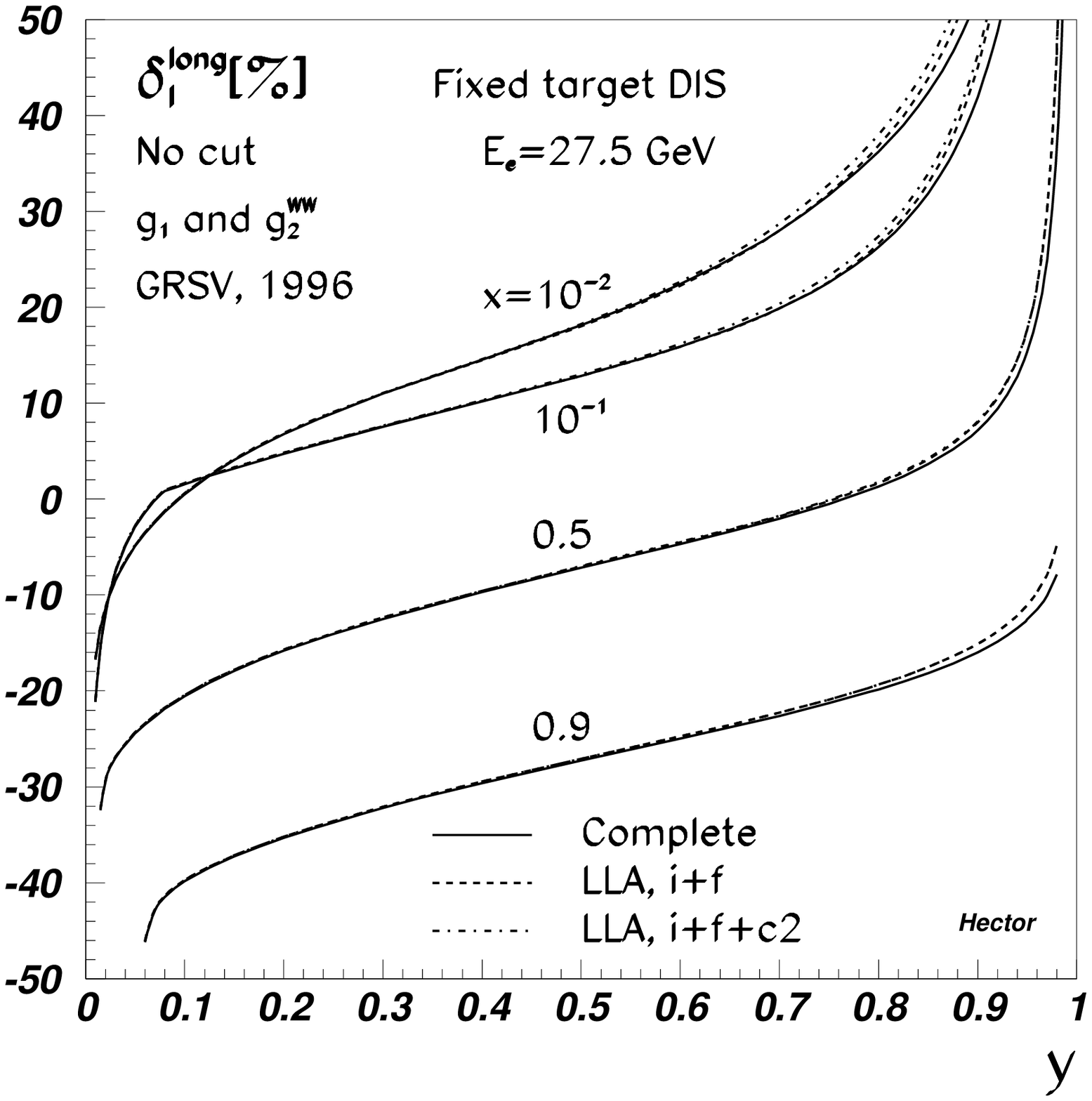,height=18cm,width=16cm}}

\vspace{2mm}
\noindent
\small
\end{center}
{\sf
Figure~5~:~$O(\alpha)$ leptonic QED correction, eq.~(\ref{COR1}), to the
polarized part of the
differential deep-inelastic scattering cross section for longitudinally
polarized protons at $\sqrt{S} = 7.4~\GeV$. Full lines~: complete
corrections; dashed lines~: initial and final-state Bremsstrahlung
contributions in LLA; dash-dotted lines~: complete LLA contributions,
eq.~(\ref{LLA1}).
}
\normalsize
%
\newpage
\begin{center}

\mbox{\epsfig{file=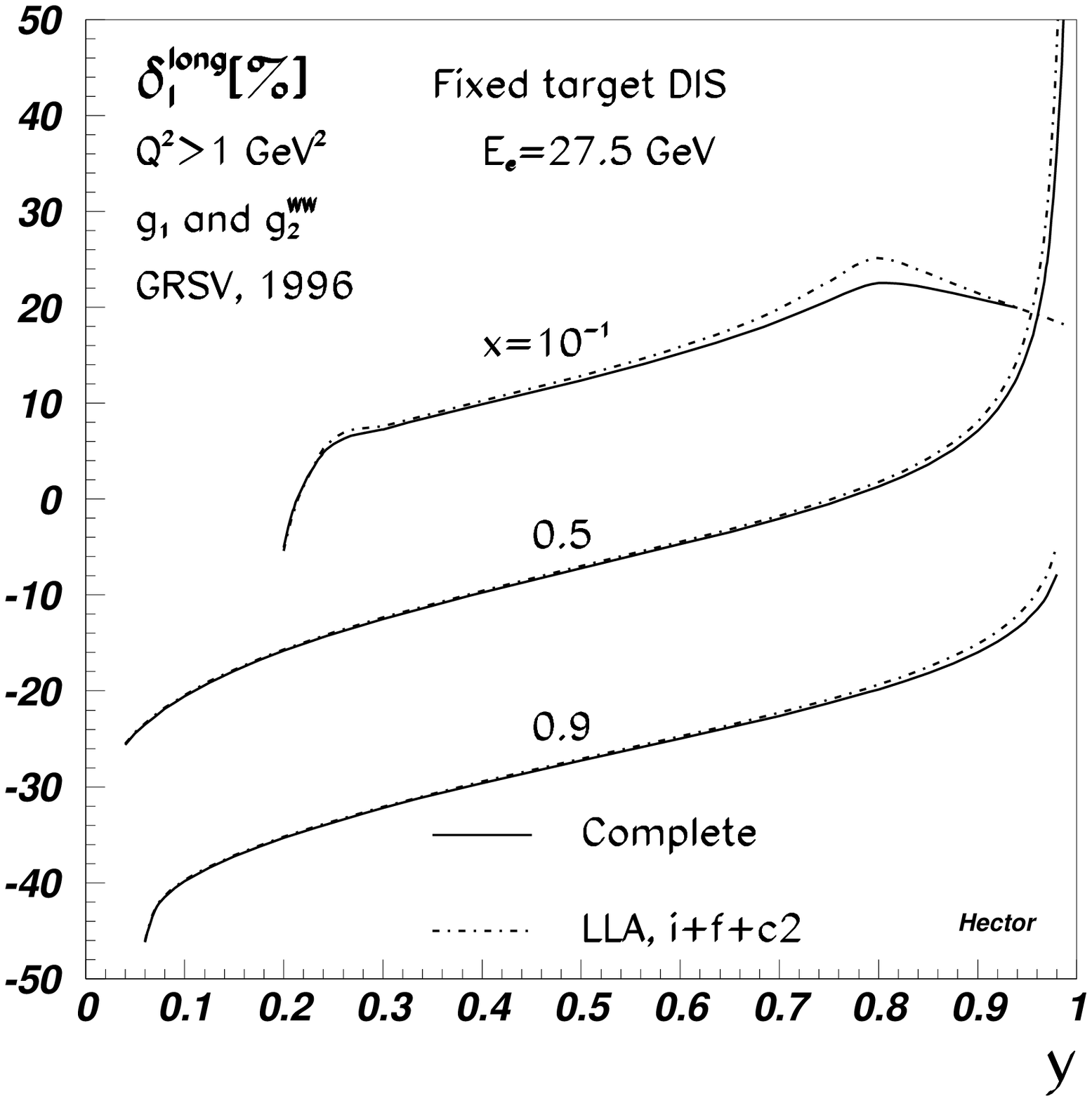,height=18cm,width=16cm}}

\vspace{2mm}
\noindent
\small
\end{center}
{\sf
Figure~6~: The same as in figure~5, but for a $Q^2$-cut of
$Q^2_h > 1 \GeV$.
Full lines~: complete corrections; dash--dotted lines~: complete
LLA corrections.
}
\normalsize
%
\newpage
\begin{center}

\mbox{\epsfig{file=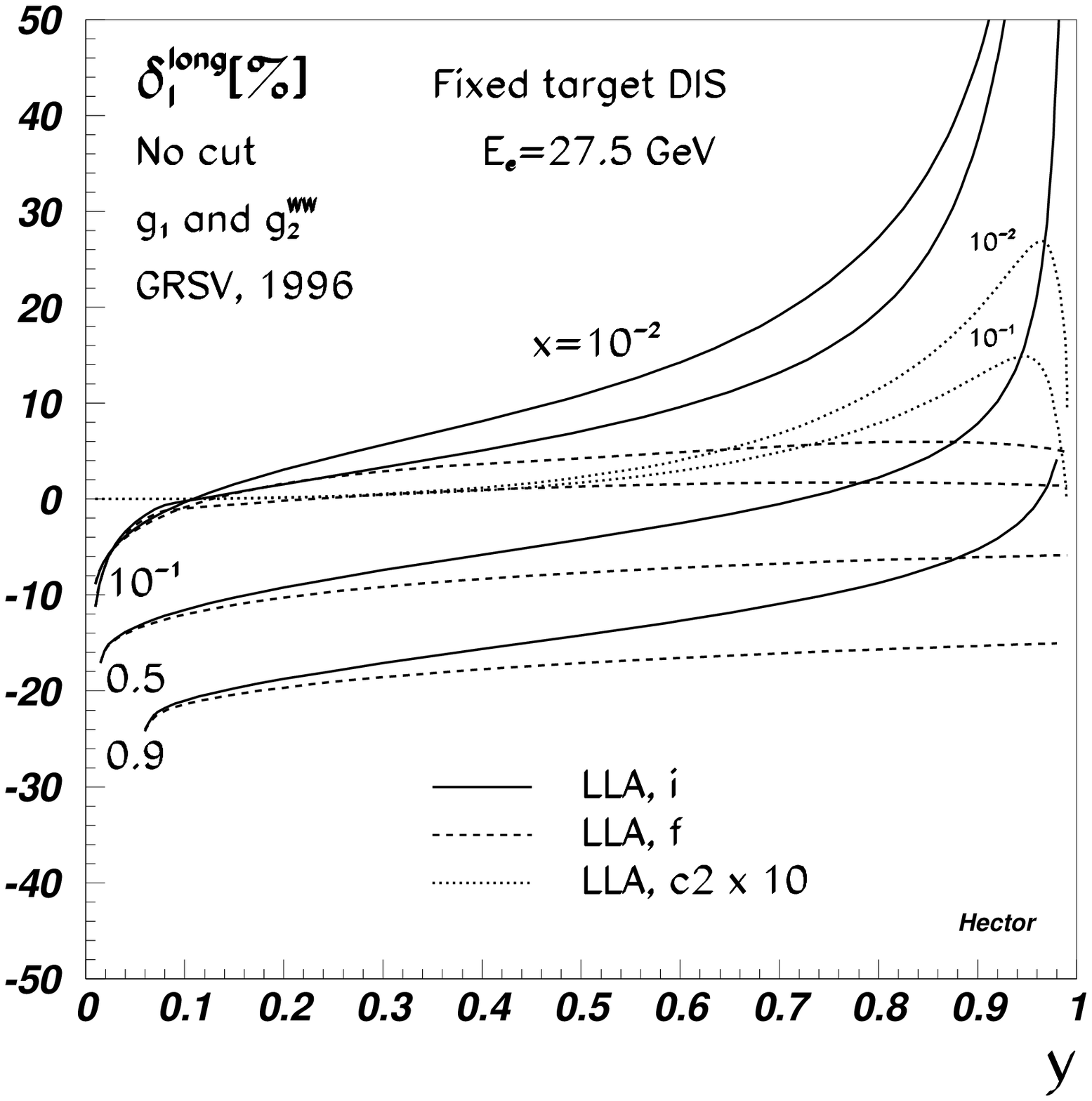,height=18cm,width=16cm}}

\vspace{2mm}
\noindent
\small
\end{center}
{\sf
Figure~7~:~Comparison of the different contributions to the
$O(\alpha)$ leptonic QED corrections in LLA
for longitudinally polarized protons at $\sqrt{S} = 7.4~\GeV$.
Full lines~: initial state radiation; dashed lines~:
final state radiation; dotted lines~: Compton
contribution,~eq.~(\ref{compt}), scaled by a factor 10.
}
\normalsize
%
\newpage
\begin{center}

\mbox{\epsfig{file=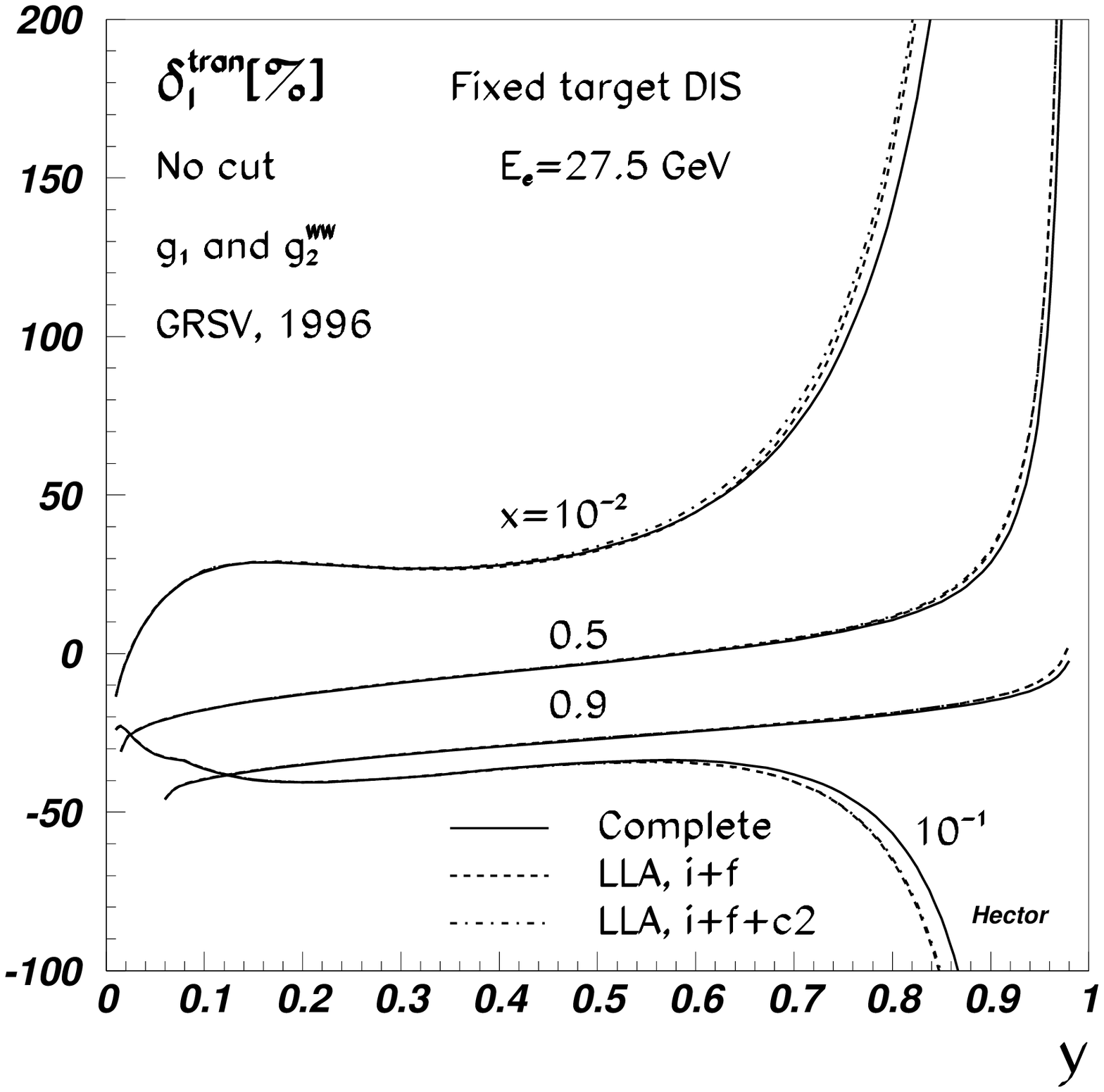,height=18cm,width=16cm}}

\vspace{2mm}
\noindent
\small
\end{center}
{\sf
Figure~8~:~$O(\alpha)$ leptonic QED correction, eq.~(\ref{COR1}), to the
polarized part of the
differential deep-inelastic scattering cross section for transversely
polarized protons at $\sqrt{S} = 7.4~\GeV$. Full lines~: complete
corrections; dashed lines~: initial and final-state Bremsstrahlung
contributions in LLA; dash-dotted lines~: complete LLA contributions,
eq.~(\ref{LLA1}).
}
\normalsize
%
\newpage
\begin{center}

\mbox{\epsfig{file=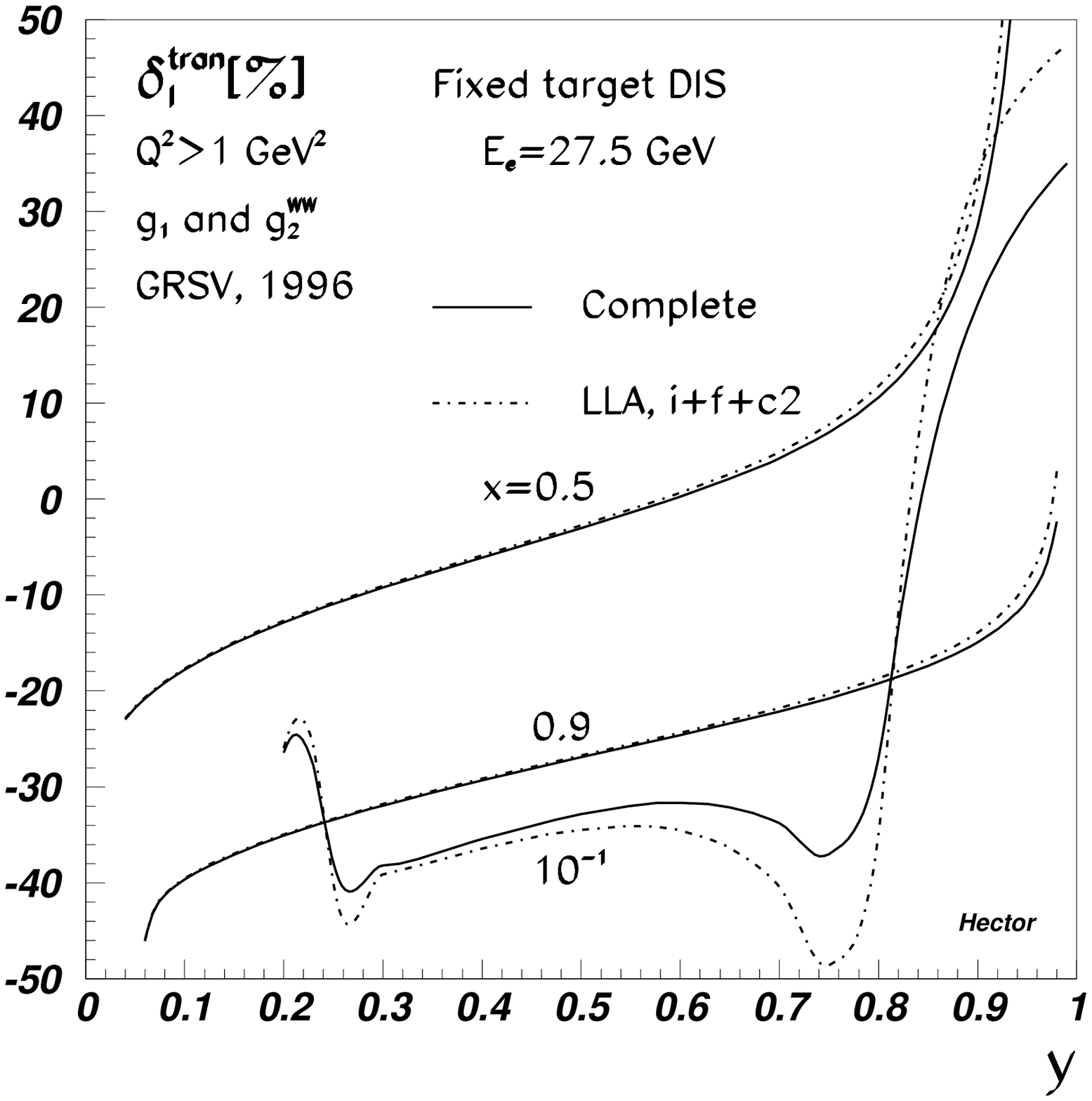,height=18cm,width=16cm}}

\vspace{2mm}
\noindent
\small
\end{center}
{\sf
Figure~9~: The same as in figure~8 applying
a $Q^2$-cut of $Q^2_h > 1 \GeV$.
Full lines~: complete corrections; dash--dotted lines~: complete
LLA corrections.
}
\normalsize
%
\newpage
\begin{center}

\mbox{\epsfig{file=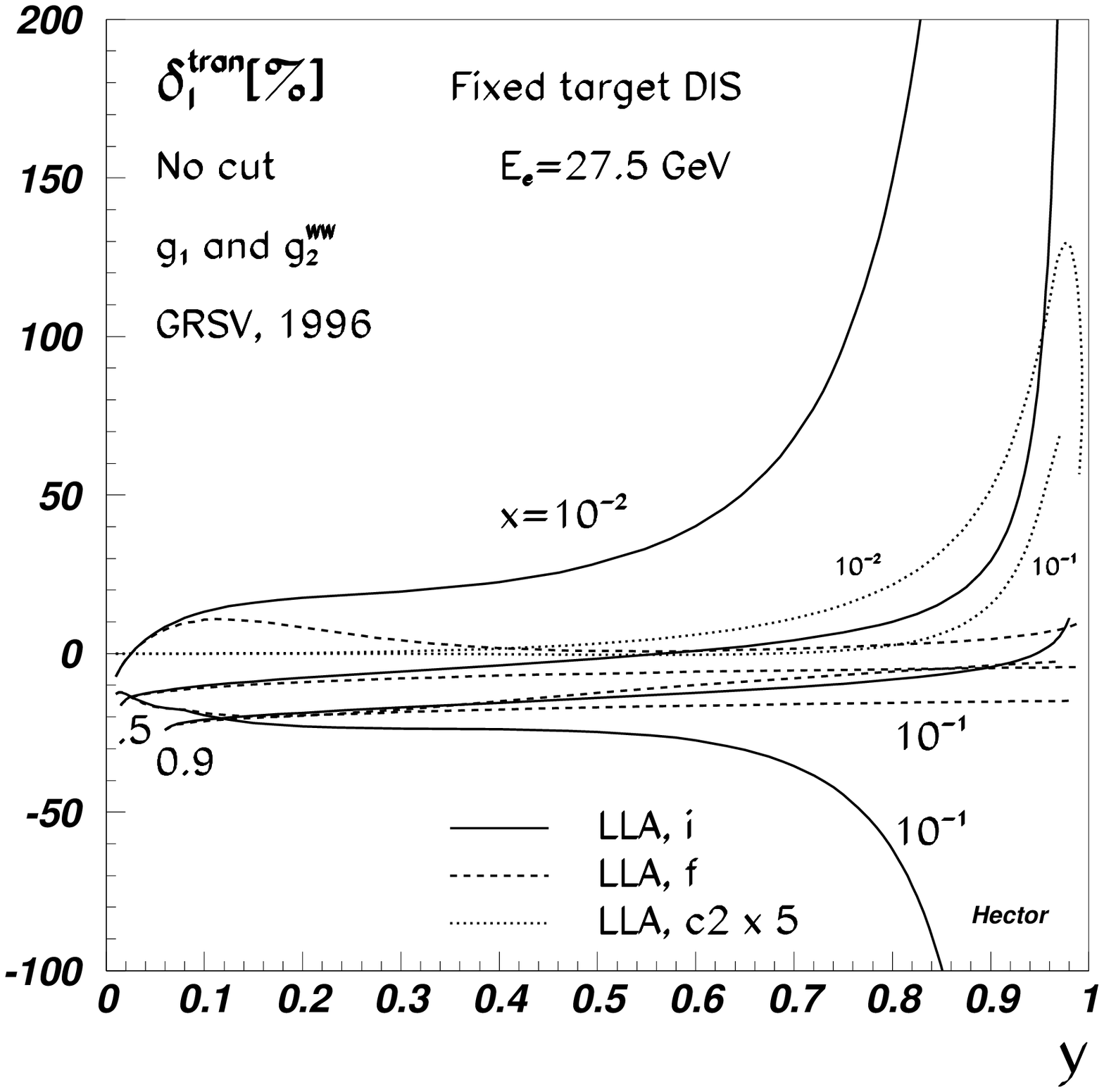,height=18cm,width=16cm}}

\vspace{2mm}
\noindent
\small
\end{center}
{\sf
Figure~10~:~Comparison of the different contributions to the
$O(\alpha)$ leptonic QED corrections in LLA
for transversely polarized protons at $\sqrt{S} = 7.4~\GeV$.
Full lines~: initial state radiation; dashed lines~:
final state radiation; dotted lines~: Compton
contribution,~eq.~(\ref{compt}), scaled by a factor 5.
}
\normalsize
%
\newpage
\begin{center}

\mbox{\epsfig{file=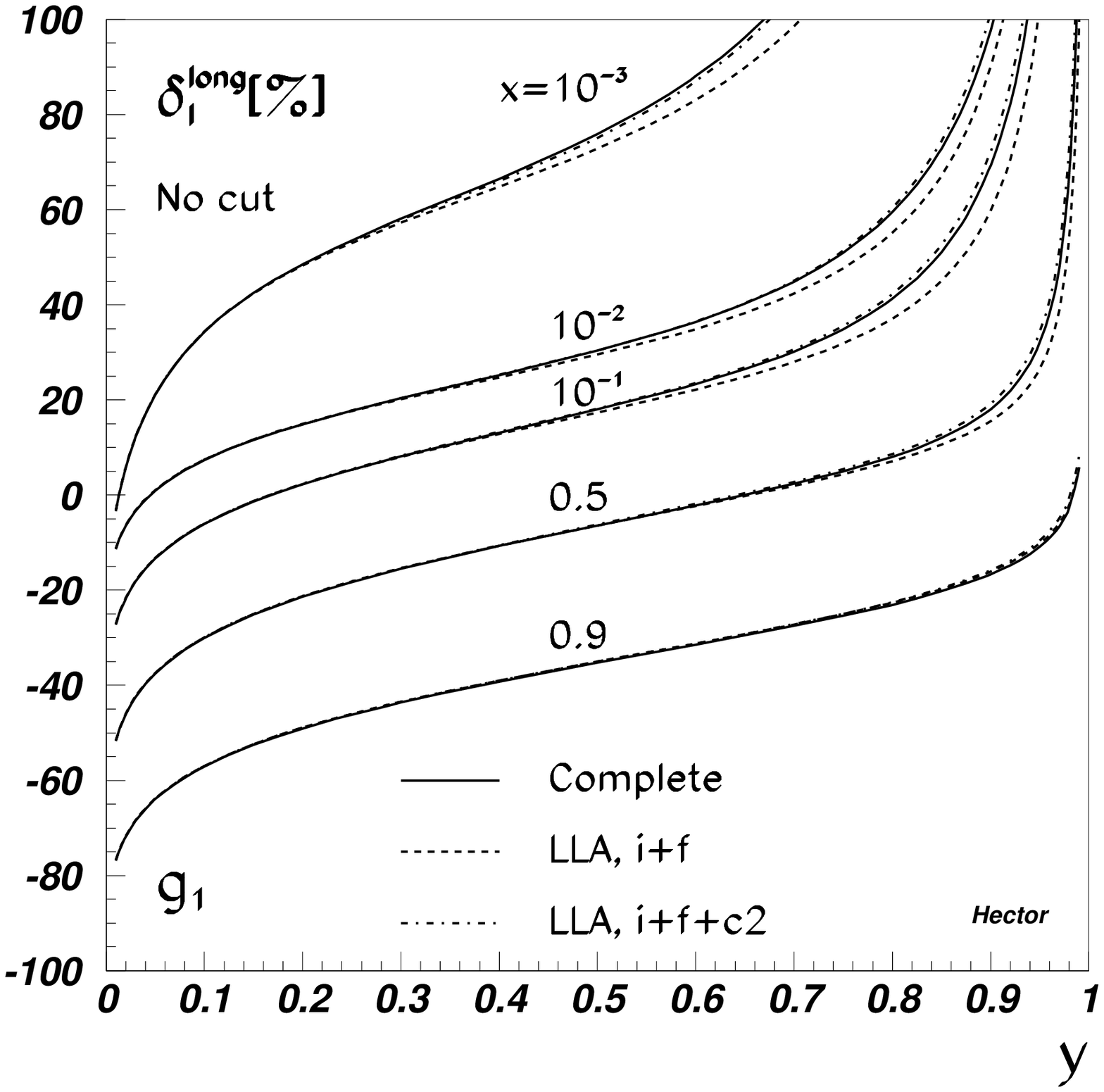,height=18cm,width=16cm}}

\vspace{2mm}
\noindent
\small
\end{center}
{\sf
Figure~11~:~$O(\alpha)$ leptonic QED correction, eq.~(\ref{COR1}), to the
polarized part of the
differential deep-inelastic scattering cross section for longitudinally
polarized protons at $\sqrt{S} = 314~\GeV$. Full lines~: complete
corrections; dashed lines~: LLA terms, eq.~(\ref{LLA1}).
\normalsize
%
\newpage
\begin{center}

\mbox{\epsfig{file=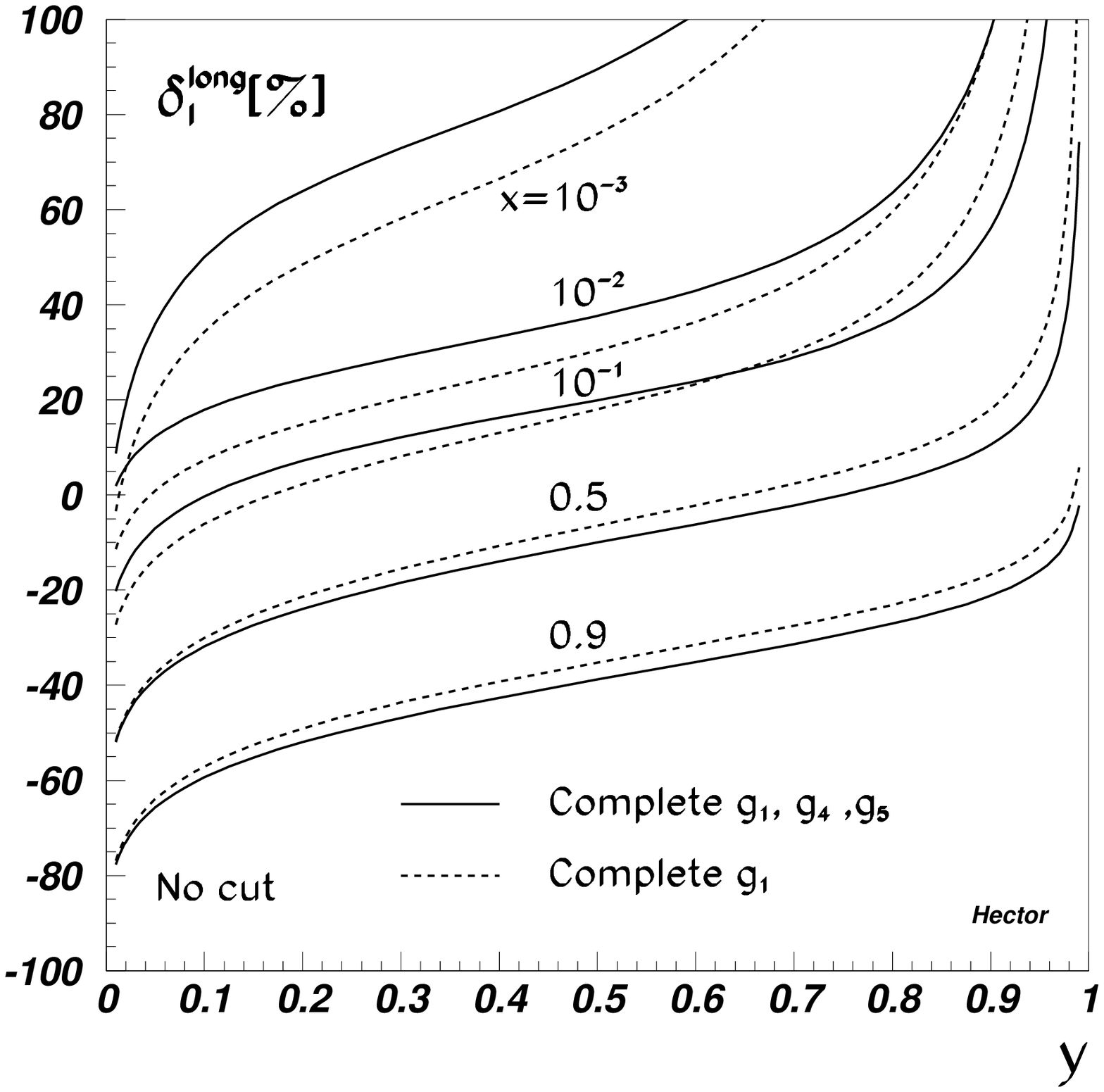,height=18cm,width=16cm}}

\vspace{2mm}
\noindent
\small
\end{center}
{\sf
Figure~12~:~$O(\alpha)$ leptonic QED correction, eq.~(\ref{COR1}), to the
polarized part of the
differential deep-inelastic scattering cross section for longitudinally
polarized protons at $\sqrt{S} = 314~\GeV$. Dashed lines~:
$\delta_1^{\rm long}$ for only the structure function $g_1$;
full lines~:  complete correction. The contributions
due to the structure functions $g_2$ and $g_3$ are of $O(M^2/S)$ and
are
not included.
}
\normalsize
\end{document}